\documentclass[acmsmall]{acmart}

\usepackage{adjustbox}
\usepackage{algorithm}
\usepackage{algorithmicx}
\usepackage{algpseudocode}
\usepackage{amsmath}
\usepackage{array}
\usepackage{balance}
\usepackage{booktabs}
\usepackage[skip=1pt,labelfont=bf]{caption}
\usepackage{calc}
\usepackage{calligra}
\usepackage{color}
\usepackage{colortbl}
\usepackage{courier}
\usepackage{csvsimple}
\usepackage{enumitem}
\usepackage{fancybox}
\usepackage{fontenc}
\usepackage{fontawesome5}
\usepackage{graphicx}
\usepackage{listings}
\usepackage{longtable}
\usepackage{lscape}
\usepackage{makecell}
\usepackage{marvosym}
\usepackage{moreverb}
\usepackage{multicol}
\usepackage{multirow}
\usepackage{pifont}
\usepackage{pgfplots}
\usepackage{rotating}
\usepackage{setspace}
\usepackage{siunitx}
\usepackage{soul}
\usepackage{subcaption}
\usepackage{tablefootnote}
\usepackage[most]{tcolorbox}
\usepackage{threeparttable}
\usepackage{tikz}
\usepackage[normalem]{ulem}
\usepackage{url}
\usepackage{wasysym}
\usepackage{xspace}

\usepgfplotslibrary{statistics}
\usetikzlibrary{matrix, shapes.geometric, arrows}
\usetikzlibrary{shapes, arrows, positioning}

\algnewcommand\algorithmicforeach{\textbf{for each}}
\algdef{S}[FOR]{ForEach}[1]{\algorithmicforeach\ #1\ \algorithmicdo}

\newcolumntype{L}[1]{>{\raggedright\let\newline\\\arraybackslash\hspace{0pt}}m{#1}}
\newcolumntype{C}[1]{>{\centering\let\newline\\\arraybackslash\hspace{0pt}}m{#1}}
\newcolumntype{R}[1]{>{\raggedleft\let\newline\\\arraybackslash\hspace{0pt}}m{#1}}

\definecolor{codegreen}{rgb}{0,0.6,0}
\definecolor{codered}{rgb}{1,0,0}
\definecolor{codegray}{rgb}{0.5,0.5,0.5}
\definecolor{codepurple}{rgb}{0.58,0,0.82}
\definecolor{backcolour}{rgb}{0.95,0.95,0.92}
\definecolor{lightgray}{gray}{0.9}

\newboolean{showcomments}
\setboolean{showcomments}{true}
\ifthenelse{\boolean{showcomments}}
 { \newcommand{\mynote}[2]{
      \fbox{\bfseries\sffamily\scriptsize#1}
        {\small$\blacktriangleright$\textsf{\emph{#2}}$\blacktriangleleft$}}}
        { \newcommand{\mynote}[2]{}}
        
\definecolor{DarkOrange}{rgb}{0.8,0.3,0.0}
\definecolor{DarkCyan}{rgb}{0.0, 0.55, 0.55}
\definecolor{DarkCyel}{rgb}{1.0, 0.49, 0.0}
\definecolor{yellow-green}{rgb}{0.6, 0.8, 0.2}

\newcolumntype{?}{!{\vrule width 1pt}}

\newcommand{\etal}{\emph{et~al.}\xspace}
\newcommand{\dataset}{\textsc{TutorLLMCode}\xspace}
\newcommand{\stdft}{\textsc{StdFT}\xspace}
\newcommand{\tool}{\textsc{MORepair}\xspace}
\newcommand*{\ie}{i.e., }

\newcommand{\find}[1]{
\begin{tcolorbox}[leftrule=1mm,toprule=0mm,bottomrule=0mm,left=1pt,right=2pt,top=2pt,bottom=2pt]
#1
\end{tcolorbox}
}

\lstdefinelanguage{mymarkdown}{
    morekeywords={*,\#, \#\#, \#\#\#},
    sensitive=false,
    morecomment=[l]{//},
    morestring=[b]",
    commentstyle=\color{codegreen},
    keywordstyle=\color{magenta},
    numberstyle=\tiny\color{codegray},
    stringstyle=\color{codepurple},
    basicstyle=\small,
    breakatwhitespace=false,         
    breaklines=true,
    breakindent=0pt,
    keepspaces=true,                 
    numbers=left,                    
    numbersep=5pt,                  
    showspaces=false,                
    showstringspaces=false,
    showtabs=false,                  
    tabsize=2,
}

\lstdefinestyle{mystyle}{
    commentstyle=\color{codegreen},
    keywordstyle=\color{magenta},
    numberstyle=\small\color{black},
    stringstyle=\color{codepurple},
    basicstyle=\scriptsize\ttfamily,
    breakatwhitespace=false,
    breaklines=true,
    captionpos=b,
    keepspaces=true,
    showspaces=false,
    showstringspaces=false,
    showtabs=false,
    tabsize=2
}

\lstdefinelanguage{diff}{
  morecomment=[f][\color{blue}]{@@},     
  morecomment=[f][\color{red}]-,         
  morecomment=[f][\color{codegreen}]+,       
  morecomment=[f][\color{red}]{---}, 
  morecomment=[f][\color{codegreen}]{+++},
  numberstyle=\tiny\color{codegray},
  numbers=left,                    
  numbersep=5pt,         
}

\setlist{noitemsep} 

\definecolor{darkpastelred}{rgb}{0.76, 0.23, 0.13}
\definecolor{ao(english)}{rgb}{0.0, 0.5, 0.0}

\definecolor{darkpastelred}{rgb}{0.76, 0.23, 0.13}
\definecolor{ao(english)}{rgb}{0.0, 0.5, 0.0}

\newboolean{useblue}
\setboolean{useblue}{false} 

\newcommand{\maybeblue}[1]{%
    \ifthenelse{\boolean{useblue}}%
    {\textcolor{blue}{#1}}%
    {#1}%
}

\AtBeginDocument{%
  }

\setcopyright{none}

\begin{document}

\title{\tool: Teaching LLMs to Repair Code via Multi-Objective Fine-tuning}
\author{Boyang Yang}
\authornote{Also affiliated with Jisuan Institute of Technology, Beijing JudaoYouda Network Technology Co. Ltd.}
\authornote{Co-first authors who contributed equally to this work.}
\affiliation{%
  \institution{School of Information Science and Engineering, Yanshan University}
  \country{China}}
\email{buaabarty@gmail.com}

\author{Haoye Tian}
\authornotemark[2]
\affiliation{%
  \institution{School of Computing and Information Systems, University of Melbourne}
  \country{Australia}
}
\email{tianhaoyemail@gmail.com}

\author{Jiadong Ren}
\affiliation{%
 \institution{School of Information Science and Engineering, Yanshan University}
 \country{China}
}
\email{jdren@ysu.edu.cn}

\author{Hongyu Zhang}
\affiliation{%
  \institution{School of Big Data and Software Engineering, Chongqing University}
  \country{China}
}
\email{hongyujohn@gmail.com}

\author{Jacques Klein}
\affiliation{%
  \institution{SnT, University of Luxembourg}
  \country{Luxembourg}}
\email{jacques.klein@uni.lu}

\author{Tegawendé F. Bissyandé}
\affiliation{%
  \institution{SnT, University of Luxembourg}
  \country{Luxembourg}}
\email{tegawende.bissyande@uni.lu}

\author{Claire Le Goues}
\affiliation{%
  \institution{School of Computer Science, Carnegie Mellon University}
  \country{USA}}
\email{clegoues@cs.cmu.edu}

\author{Shunfu Jin}
\authornote{Corresponding author.}
\affiliation{%
 \institution{School of Information Science and Engineering, Yanshan University}
 \country{China}
}
\email{jsf@ysu.edu.cn}

\renewcommand{\shortauthors}{Yang et al.}

\begin{abstract}

Within the realm of software engineering, specialized tasks on code, such as program repair, present unique challenges, necessitating fine-tuning Large language models~(LLMs) to unlock state-of-the-art performance. Fine-tuning approaches proposed in the literature for LLMs on program repair tasks generally overlook the need to reason about the logic behind code changes, beyond syntactic patterns in the data. High-performing fine-tuning experiments also usually come at very high computational costs. With \tool, we propose a novel perspective on the learning focus of LLM fine-tuning for program repair: we not only adapt the LLM parameters to the syntactic nuances of the task of code transformation (objective \ding{182}), but we also specifically fine-tune the LLM with respect to the logical reason behind the code change in the training data (objective \ding{183}). Such a multi-objective fine-tuning will instruct LLMs to generate {\em high-quality} patches.

We apply \tool to fine-tune four open-source LLMs with different sizes and architectures. Experimental results on function-level and repository-level repair benchmarks show that the implemented fine-tuning effectively boosts LLM repair performance by 11.4\% to 56.0\%. We further show that our fine-tuning strategy yields superior performance compared to the state-of-the-art approaches, including standard fine-tuning, Fine-tune-CoT, and RepairLLaMA.

\end{abstract}

\maketitle

\section{Introduction}
Large language models have achieved promising performance on a variety of tasks in different domains. 
In software engineering, automated program repair (APR) is one of many code-related tasks that has greatly benefited from the general knowledge encoded in prominent models such as GPT-4~\cite{achiam2023gpt4}. 
Recent studies~\cite{xia2023keep,fan2022improving,jiang@impact} have indeed shown that LLMs can even outperform traditional APR tools. Researchers have achieved these results using two main strategies: prompt engineering and fine-tuning. Indeed, to steer LLMs towards adapting to the specific format of repair, few-shot learning techniques have been employed~\cite{cao2023study,phung2023generating,peng2024domain,joshi2023repair,feng2024prompting} where a small set of example patches are provided in the prompt along with the buggy code to repair. While few-shot-based approaches have shown better performance than initial zero-shot-based attempts~\cite{fan2023automated,prenner2022can,fu2022vulrepair},  prompting is inherently limited by the pre-trained model capabilities. Furthermore, prompting often fails to produce high-quality patches within the constraints of developers' attempt limits~\cite{noller2022trust}. In contrast, fine-tuning-based approaches~\cite{jiang@impact,berabi2021tfix,lajko2022fine,huang2023empirical,silva2023repairLlama} strive to refine the fundamental capabilities of LLMs and have therefore demonstrated substantially greater potential in achieving reliable program repair. In practice, fine-tuning consists of adapting a pre-trained LLM on a very specific dataset, such as patches, or task, such as program repair, enabling the model to refine its knowledge and improve performance in targeted areas~\cite{jiang2021cure}. Unfortunately, the existing literature proposes approaches that still face two major limitations: 

\begin{itemize}[leftmargin=*]
\item[\ding{172}] \textbf{Need for Reasoning on Repair Logic}: The program repair task is complex: it demands some deep comprehension of control and data flow of the developer's intentions in the design of the buggy code, and finally of the intrinsic repair logic. Yet, most of the standard fine-tuning approaches for LLM-based program repair focus on optimizing the training dataset~\cite{jiang@impact,lajko2022fine,2020t5}. Thus, while with such approaches, the LLMs can be refined to notice some repair patterns, the actual logical reasoning behind the repair operation ("the why") is not explicitly learned.

\item[\ding{173}] \textbf{High Cost}: Fine-tuning for program repair generally requires large datasets to achieve state-of-the-art performance. In recent works, Lajk{\'o}~\etal~\cite{lajko2022fine} used 16k samples to fine-tune GPT-2, while  RepairLLaMA~\cite{silva2023repairLlama} was fine-tuned with about 30-50k code pairs. Such large datasets further suggest a significant expenditure of computational resources. Creating and expanding these datasets takes substantial effort and time, emphasizing the resource-intensive nature of fine-tuning in program repair. With limited public datasets available, manual construction of training data further increases labor costs.

\end{itemize}

\textbf{This paper.} To address these limitations, we propose a new fine-tuning objective using natural language explanations of code changes to capture the logic underlying a given repair operation. This objective, which seeks conversational guidance, is considered in addition to the classical objective of learning code transformations. \tool is thus designed as a novel, effective program repair framework leveraging multi-objective fine-tuning for LLM-based program repair.

By focusing on conversational guidance, i.e., natural language, \tool ensures that the learning is programming language-independent, making it suitable for multilingual repair scenarios. Furthermore, by conducting multi-objective learning, we indirectly scale up the learning dataset: more pattern combinations can be explored in a small dataset. We also observe that conversational guidance presents the benefit of providing various potential fix strategies that extend beyond the confines of a specific buggy code. As such, our approach does not depend on large-scale datasets for fine-tuning that are required by prior works. Experimentally, we show that with an order of magnitude smaller dataset, we achieve higher fine-tuning performance than prior works. Finally, to account for insufficient/missing patch descriptions, we rely on LLMs to generate high-quality patch guidance. The successful application of such automatically generated guidance is essential as it relieves APR from this expensive human input~\cite{tang2023just}.

We apply \tool to fine-tune four open-source LLMs, namely CodeLlama-13B-instruct~\cite{roziere2023code}, CodeLlama-7B-instruct~\cite{roziere2023code}, StarChat-alpha~\cite{Tunstall2023starchat-alpha}, and Mistral-Instruct-7B-v0.1~\cite{jiang2023mistral}, which are chosen to represent a variety of model sizes and architectures. These are assessed against two new repair benchmarks, EvalRepair-C++ and EvalRepair-Java, which we produced based on HumanEval~\cite{chen2021humaneval} by including augmented test cases to avoid patch overfitting~\cite{yang2017better}. The experiments demonstrate that the proposed fine-tuning technique effectively improves the LLM performance on the program repair task: CodeLlama-13B-instruct performance is improved by 18.8\% and 11.4\% on the EvalRepair-C++ and EvalRepair-Java benchmarks, respectively. Similar performance improvements have been observed across all LLMs. We also show that \tool is indeed superior to the fine-tuning approaches used for state-of-the-art models such as Fine-tune-CoT~\cite{ho2022large} and RepairLLaMA~\cite{silva2023repairLlama}. Finally, we show that \tool has the ability to narrow the performance gap between small open-source models and larger closed-source models. 

The main contributions of our work are as follows:

\begin{itemize}[noitemsep,topsep=0pt,leftmargin=2em]
\item {\bf Approach.} We propose {\tool}, a novel program repair framework leveraging multi-objective fine-tuning specifically for LLM-based program repair. \tool helps LLMs adapt a precise understanding of the reasoning logic behind the repair process, thereby enabling them to generate high-quality patches.

\item {\bf Benchmarks.} 
We provide two new repair benchmarks, EvalRepair-C++ and EvalRepair-Java, consisting of 164 and 163 patches (pairs of code samples), respectively. EvalRepair-C++ was created by manually introducing bugs into the ground truth C++ code from HumanEval-X~\cite{zheng2023codegeex}, while EvalRepair-Java is derived from HumanEval-Java~\cite{jiang@impact}. To mitigate patch overfitting impact on the reported performance metrics, we augment the original test suites of both benchmarks: we indeed observe a decline of up to $\sim$9 percentage points in terms of Top-10 repair predictions from CodeLlama-13B-instruct when we apply it to the new EvalRepair-C++ benchmark. Additionally, we provide two repository-level benchmarks: D4J-Repair (371 Java bugs) and SWE-Repair (204 Python bugs), curated subsets of Defects4J and SWE-Bench, respectively.

\item {\bf Experiments.} We conduct a comprehensive evaluation of \tool's effectiveness for improving the performance of open source LLMs, as well as its generalizability across various LLMs and different programming languages. The assessment also considers baseline models and baseline fine-tuning approaches. 

\item {\bf Insights.} Ablation studies confirm that optimizing only for patch correctness (\stdft) or patch explanation (Fine-tune-CoT) is suboptimal, while jointly optimizing both achieves the best repair accuracy, highlighting the necessity of multi-objective fine-tuning. We further find that the value of LLM-generated guidance extends beyond length—it provides richer insights into the underlying bug, leading to more logical repairs. Finally, \tool consistently outperforms other state-of-the-art fine-tuning methods, validating its effectiveness.

\end{itemize}

Our research artifacts, including code and the reproduction data, are publicly available at:

\begin{center}
\href{https://github.com/buaabarty/morepair}{\color{blue}https://github.com/buaabarty/morepair}
\end{center}

\section{Motivating Example}
\label{motivating}

\begin{figure}[h]
\centering
\begin{minipage}{.8\textwidth}
\centering
\captionsetup{justification=centering}
\begin{lstlisting}[language=diff,numbers=none]
--- 0_58980_RE_SEGV.cpp
+++ 0_58980_AC.cpp
@@ -35,8 +35,8 @@
                  num.push(cal(tmp1,tmp2,opr));
              }
              op.pop();
          }else if (s[i] == '+' || s[i] == '-') {
-            while (!op.empty()) {
+            while (!op.empty()&&op.top()!='(') {
                 int tmp1 = num.top();
                 num.pop();
                 int tmp2 = num.top();
\end{lstlisting}
\end{minipage}
\caption{An example patch from \dataset.}
\label{fig:buggycode_dataset}
\end{figure}

Figure~\ref{fig:buggycode_dataset} provides an example patch for repairing a C++ program in the \dataset dataset. The bug is related to handling precedence in the operator of arithmetic expressions. When operation expressions are mixed or rather interleaved, the buggy program could yield an inaccurate evaluation of the arithmetic expressions. In \dataset, this repair case is an example of a patch that was necessary to fix the solution provided by a student to an \textit{Expression Evaluation} problem. The solution to this engineering problem is to introduce parentheses in basic arithmetic operations (addition, subtraction, multiplication, division). Writing this code without taking into account the presence of parentheses may lead to bugs. 

\begin{lstlisting}[language=mymarkdown,frame=tb,basicstyle=\footnotesize, label={lst:cot_human}, caption= Human-generated guidance for yielding the patch in Figure~\ref{fig:buggycode_dataset},numbers=none]
Line 39, also ensure that the top element of the stack is not a left parenthesis.
\end{lstlisting}

The human-generated guidance in Listing~\ref{lst:cot_human} accurately points to the buggy location, and also conceptually describes the necessary code checks that are missing. We postulate that such information is valuable as it reveals the logical reasoning behind the requested repair and abstracts away from the particular syntax of the associated program, potentially enabling a fine-tuning process with this information to generalize to a broader set of programs.

Listing~\ref{lst:cot_llm} depicts the LLM-generated guidance for this repair case. The LLM was prompted with the buggy along with the fixed code and instructed to provide step-by-step descriptions of the repair logic. On average, the human-written explanations are 36 words, while the LLM explanations are 337 words, providing significantly more detail and multiple potential fix strategies. Interestingly, the LLM-generated guidance goes into more depth, offers a more comprehensive and contextual analysis of the buggy code, and is typically quite structured. 
The LLM identifies the primary issue: correctly handling of '+' and '-' operators when not preceded by a '(', highlighting four potential problems. Concretely, the LLM-generated guidance outlines in detail some corrections by refining the logic for operator precedence, ensuring accurate parenthesis handling, mentioning parameter order adjustments in the \textit{cal} function for subtraction and division, and introducing error handling for enhanced robustness.

Given that LLMs often benefit when learning from more structured input, we hypothesize that the detailed and structured nature of the LLM-generated explanations, in addition to their greater length and multiple fix strategies, may provide a richer learning signal for LLMs during fine-tuning compared to the human-written explanations. This could potentially lead to improved generalization to a wider range of bugs.

\begin{lstlisting}[language=mymarkdown,frame=tb,basicstyle=\footnotesize,label={lst:cot_llm}, caption= LLM-generated guidance for the patch in Figure~\ref{fig:buggycode_dataset},numbers=none]
Certainly! Let's go through the code step by step and identify the issues that need to be fixed:
1. **Operator Precedence**: In the original code, when encountering a '+' or '-', it immediately evaluates all operations in the stack without considering operator precedence. This is incorrect because '*' and '/' should have higher precedence and should be evaluated first.
   **Fix**: When encountering a '+' or '-', only evaluate the top of the stack if it is not a '(' and if the operator at the top of the stack is '+' or '-'. Do not evaluate '*' or '/' immediately; they should be evaluated when the next '+' or '-' is encountered or at the end of the expression.
2. **Parentheses Handling**: The original code correctly handles the opening parenthesis '(', but when it encounters a closing parenthesis ')', it evaluates all operations until it finds the corresponding '('. This is correct, but it does not consider operator precedence within the parentheses.
   **Fix**: When encountering a ')', evaluate the operations in the stack until you reach the corresponding '('. Make sure to evaluate '*' and '/' before '+' and '-'.
3. **End of Expression**: At the end of the expression, the original code evaluates all remaining operations in the stack. This is correct, but again, it does not consider operator precedence.
   **Fix**: When reaching the end of the expression, evaluate the remaining operations in the stack while considering operator precedence. Evaluate '*' and '/' before '+' and '-'.
4. **Function `cal`**: The `cal` function has the parameters in the wrong order for subtraction and division. This will lead to incorrect results when these operations are performed.
   **Fix**: Swap the order of the parameters `x` and `y` in the subtraction and division cases within the `cal` function to ensure the correct order of operands.
5. **Error Handling**: The original code does not handle errors or invalid input. While the problem statement guarantees valid input, it's good practice to include error handling.
   **Fix**: Add error handling to check for invalid characters, mismatched parentheses, or other potential input errors. This is not explicitly required by the problem statement but is a good practice.
By addressing these issues, the code should correctly evaluate expressions with proper operator precedence and parentheses handling. Remember to test the code with various expressions to ensure that all edge cases are handled correctly.
\end{lstlisting}

Our main intuition in this work is that the conversational guidance generated from repair examples will significantly benefit a model fine-tuning procedure for program repair. 
\section{Approach}
\label{approach}
This section provides an overview of our proposed approach, followed by a detailed description of the methodology, which is divided into specific steps across several subsections.

\begin{figure}[h]
\centering
\includegraphics[width=\textwidth]{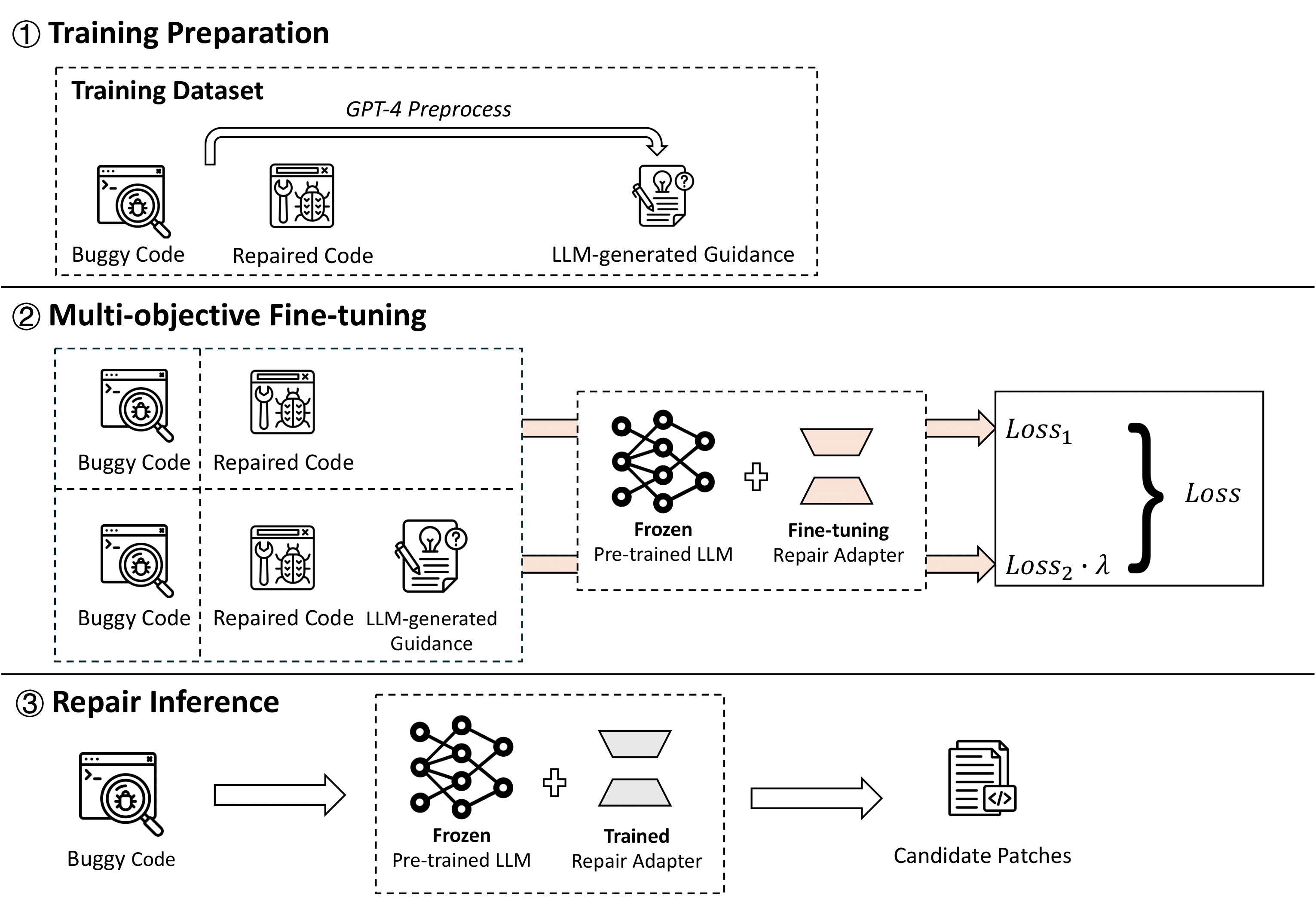}
\caption{Overview of \tool: The process unfolds in three phases—preparation, fine-tuning, and inference for code repair.}
\label{fig:process}
\end{figure}

\noindent
\textbf{[Overview]}: We introduce \tool, a novel program repair framework leveraging multi-objective fine-tuning that empowers open-source LLMs to grasp repair logic and produce high-quality patches effectively. Figure \ref{fig:process} illustrates our approach, which unfolds in three phases: training preparation, multi-objective fine-tuning, and repair inference.

During the \textbf{Training Preparation phase}, we constructed a dataset \dataset, consisting of 1,535 pairs of buggy and repaired code submitted by students from a company's online programming education platform. \dataset covers 45 different programming problems, each accompanied by a human-written rationale for the repair provided by the company's tutors, who are paid domain experts in the programming problems and solutions. This preparation includes LLM-generated guidance generated by GPT-4 to elucidate the nature of code bugs and their fixes~(as detailed in Section \ref{phase1}). The \textbf{Multi-objective Fine-tuning phase} applies the principles of multi-objective learning, targeting two specific learning objectives: (1) generating repaired code and (2) producing repaired code with guidance that explains the repair logic. Leveraging QLoRA allows for fine-tuning a low-rank adapter while freezing the original LLM parameters, cutting down the trainable parameters to only 1.88\% and thus minimizing computational costs. In the \textbf{Repair Inference phase}, the ensemble of the pre-trained LLM and the fine-tuned repair adapter generates candidate patches for the provided buggy code, whose correctness is validated through the test cases from benchmarks.

\subsection{Training Preparation}
\label{phase1}
The initial step in our approach involves preparing the training dataset for fine-tuning. We utilize a dataset \textbf{\dataset}, provided by a company, which includes 1,535 pairs of single-file C++ buggy codes submitted by students from a company's online programming education platform. \dataset covers 45 different programming tasks, each accompanied by a human-written rationale for the repair provided by the company's programming tutors, who are paid domain experts in solving and teaching programming problems. To avoid potential data leakage between training and evaluation, we carefully verified that there is no overlap between the 45 programming tasks in \dataset and the 164 tasks in HumanEval-X that were used to create our benchmarks EvalRepair-C++ and EvalRepair-Java.

While \dataset contains human-written repair rationales, our analysis shows their limitations: they average only 36 words and mainly highlight fix locations rather than repair reasoning (Listing \ref{lst:cot_human}). In contrast, GPT-4-1106-preview generates more comprehensive explanations (Listing \ref{lst:cot_llm}), detailing repair logic, step-by-step reasoning, potential fix strategies, and implementation considerations. Thus, we use GPT-4-1106-preview to generate rationales solely from buggy-fixed code pairs without relying on human-written explanations. We use greedy decoding (temperature=$0.0$) without additional sampling and set the maximum token length to 1024 to ensure deterministic and comprehensive explanation generation. Figure \ref{fig:prompt_cot} shows the prompt template used for GPT-4 to generate guidance.

We employ a two-phase process to characterize the quality of the generated guidance. Initially, we assess format compliance and analyze the structural characteristics of the explanations. Our analysis of all 1,535 GPT-4 generated rationales shows they are consistently detailed and well-structured: they contain between 139 and 454 words (averaging 337 words per rationale), with 100\% incorporating organized lists and 90.9\% utilizing emphasis markers (such as headers and bold text) to highlight key points. As a final examination step, we manually inspect a random 10\% sample of the generated guidance to assess their completeness, accuracy, and alignment with actual code changes.

To facilitate reproducibility and future research, we have made \dataset available through an authorized API at \url{https://github.com/buaabarty/morepair}.

\begin{figure}[H]
\centering
\begin{minipage}{\textwidth}
\centering
\captionsetup{justification=centering}
\begin{lstlisting}[language=mymarkdown,frame=tb,basicstyle=\footnotesize]
  This is a programming task description along with a buggy code:
  {{description}}
  {{buggy code}}
  This is a repaired code:
  {{repaired code}}
  Please think step by step and tell me how to fix the buggy code.
\end{lstlisting}
\end{minipage}
\caption{The prompt to generate guidance utilizing GPT-4.}
\label{fig:prompt_cot}
\end{figure}

\subsection{Dataset Analysis}
\label{sec:dataset_analysis}
To understand the characteristics and complexity of our training data, we conducted a comprehensive analysis of \dataset.

\begin{figure}[ht]
\centering
\includegraphics[width=.7\textwidth]{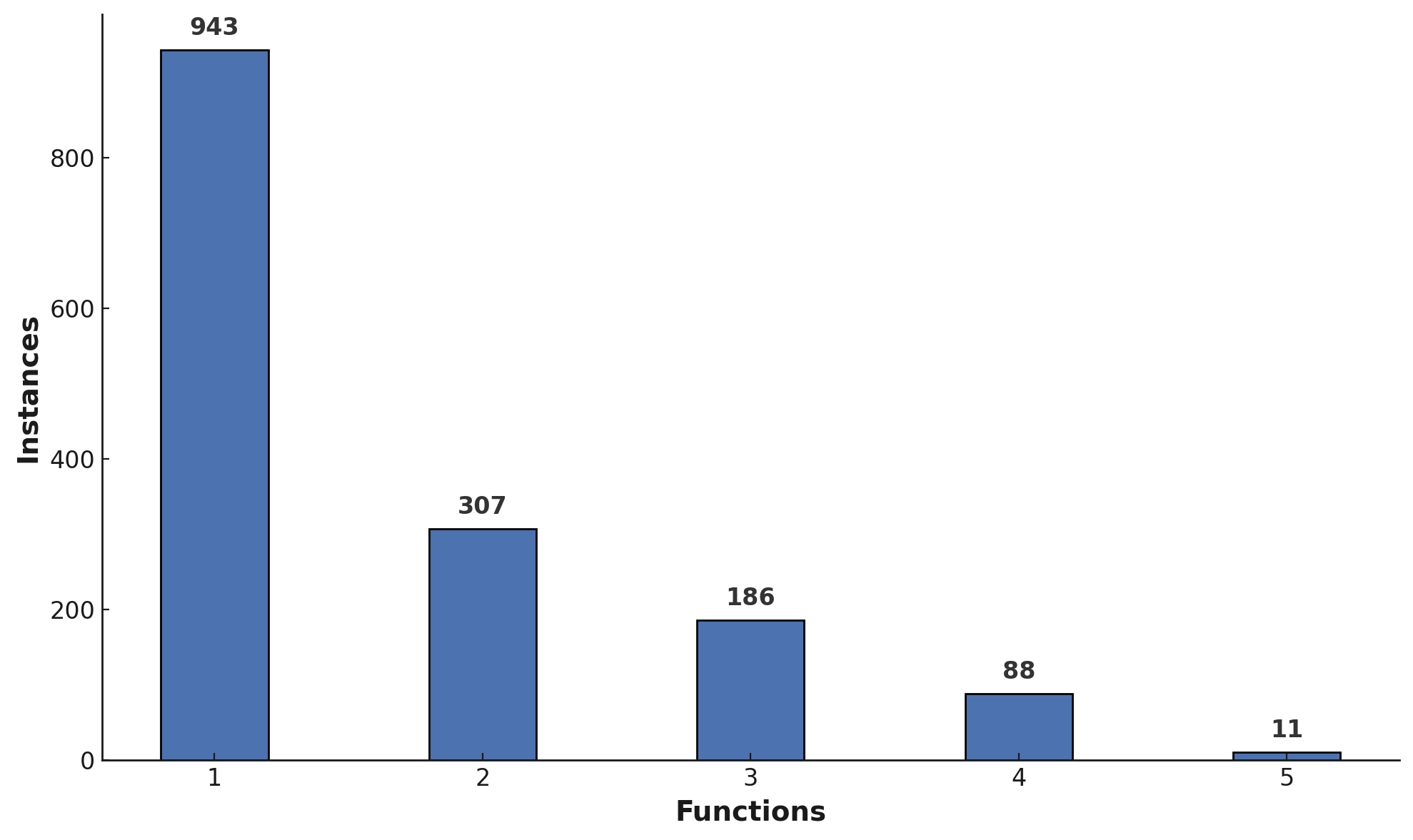}
\caption{Distribution of function count per instance in \dataset.}
\label{fig:dataset_funcs}
\end{figure}

Figure \ref{fig:dataset_funcs} presents the distribution of the number of functions in the buggy codes in \dataset. The analysis shows that while the majority (61.4\%) of the codes contain a single function, a substantial portion (38.6\%) consists of multiple functions, indicating considerable structural diversity in the dataset.

\begin{figure}[ht]
\centering
\includegraphics[width=.7\textwidth]{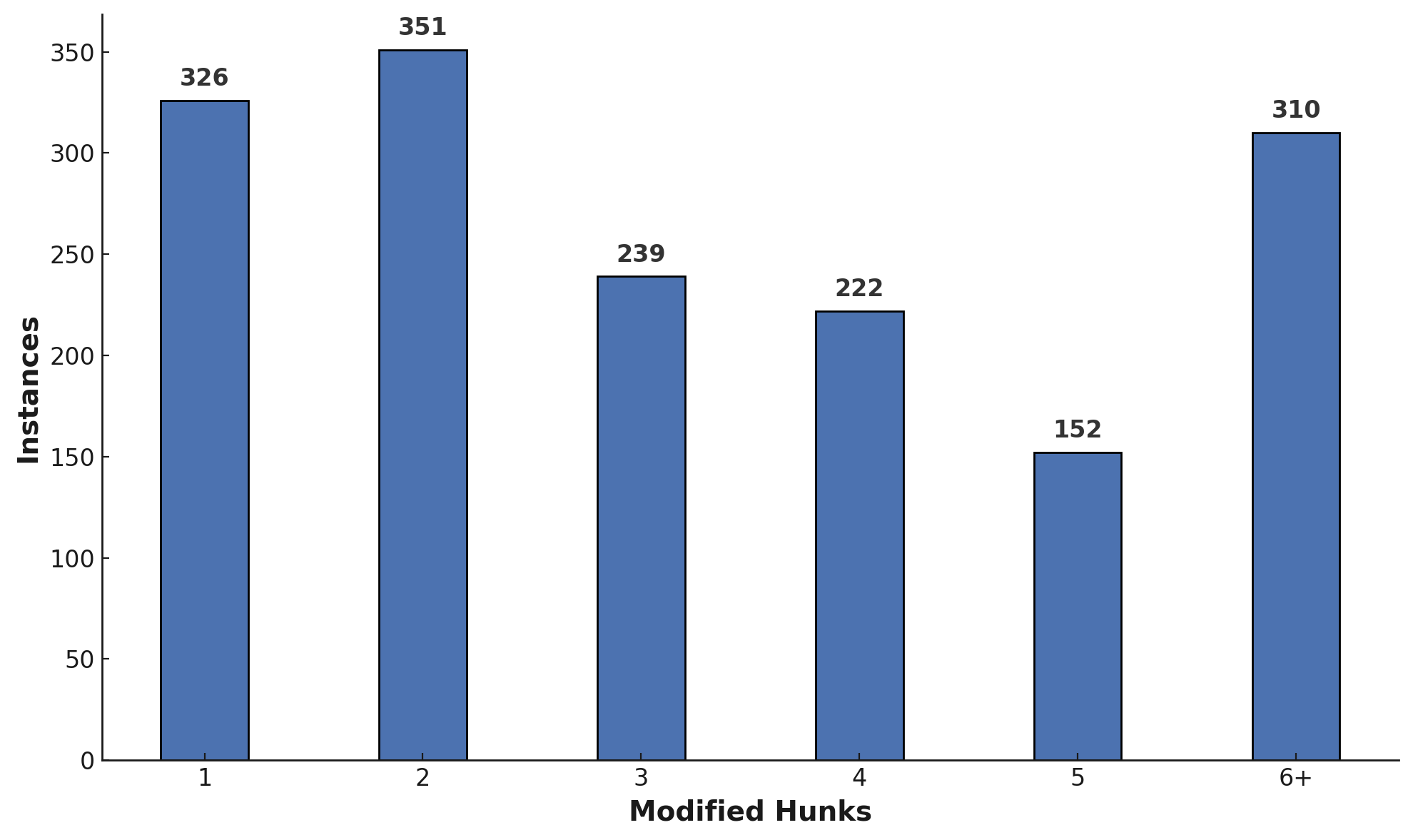}
\caption{Distribution of modified hunks count per instance in \dataset.}
\label{fig:dataset_hunks}
\end{figure}

Figure \ref{fig:dataset_hunks} shows the distribution of modified hunks, calculated as the difference between the buggy code and the student-corrected ground truth. The data reveals a clear predominance of multi-hunk modifications, with 78.8\% of repairs containing multi-hunk modifications compared to 21.2\% with single-hunk modifications.

\begin{figure}[ht]
\centering
\includegraphics[width=.7\textwidth]{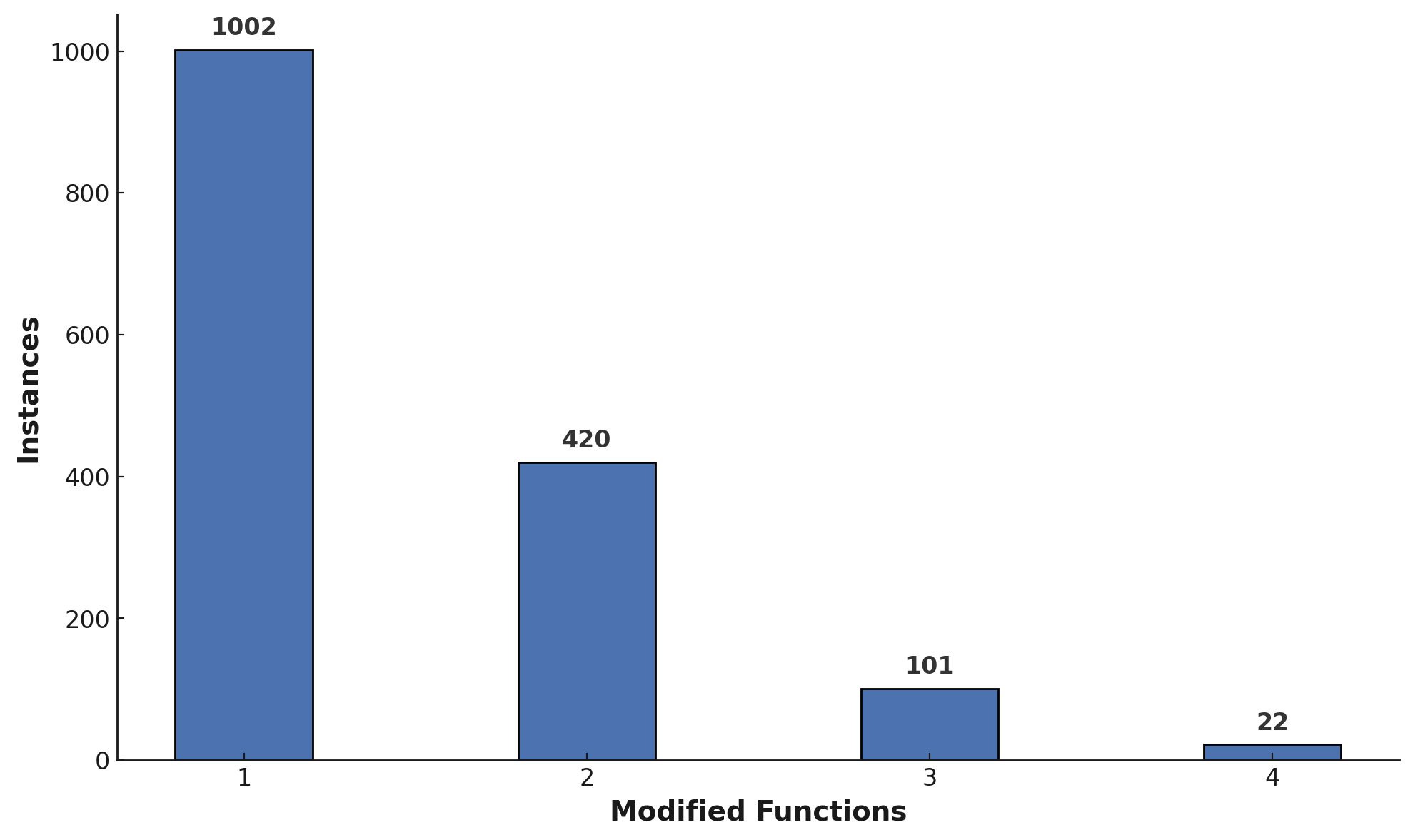}
\caption{Distribution of modified functions count per instance in \dataset.}
\label{fig:dataset_modifyfuncs}
\end{figure}

Figure \ref{fig:dataset_modifyfuncs} illustrates the distribution of modified functions in repairs. While the majority of repairs (64.6\%) involve changes to a single function, a significant proportion (35.4\%) spans multiple functions. These statistics collectively highlight the diverse complexity of repairs in \dataset.

\subsection{Multi-objective Fine-tuning}
\label{training}

The second step of our approach involves fine-tuning LLMs through multi-objective learning. Multi-objective learning, a paradigm in machine learning, aims to leverage relevant information across multiple tasks simultaneously to enhance the performance of each task~\cite{zhang2018overview}. In this context, we propose our framework \textbf{\tool}, which applies multi-objective learning to enhance LLMs' ability to repair code by optimizing patch correctness and explanatory clarity during fine-tuning. Unlike previous CoT fine-tuning approaches that concatenate code and rationale as a single objective~\cite{ho2022large}, our approach \tool enables the LLMs to generate high-quality patches for buggy code through two objectives: (1) generating repaired code and (2) producing repaired code accompanied by guidance that clarifies the nature of the bugs and their logic. To optimize for these objectives, we calculate separate losses for each, denoted as $Loss_1$ for generating the repaired code, and $Loss_2$ for producing both the repaired code and its explanatory guidance. These losses are then combined using the following equation:

\begin{equation}
\label{eq:loss}
Loss = (1-\alpha) Loss_{1} + \alpha Loss_{2}
\end{equation}

where \(\alpha\) is the weighting coefficient that balances the importance of each objective during the training process. We set $\alpha=0.5$ to maintain equal contributions from both objectives, which will be detailed in Section \ref{hyper-parameters}.

Given a buggy code sequence \(\mathbf{x} = (x_1,...,x_n)\), where each token represents a component of an erroneous function-level code snippet, our approach generates a corrected target sequence \(\mathbf{y}\). Depending on the training objective, \(\mathbf{y}\) has two possible forms:

For the first objective (\( Loss_1 \)), the target sequence consists of the repaired code:  
\[
\mathbf{y} = (y_1,...,y_m)
\]
Where \((y_1,...,y_m)\) represents the corrected version of the input code \(\mathbf{x}\).

For the second objective (\( Loss_2 \)), the target sequence contains both the repaired code and its explanation:  
\[
\mathbf{y} = (y_1,...,y_m,y_{m+1},...,y_{m+k})
\]
Where \((y_{m+1},...,y_{m+k})\) provides natural language guidance explaining the repair process. We do not introduce explicit separators between the repaired code and its explanation.

The model generates these sequences in an auto-regressive manner, predicting each token \( y_i \) based on the input sequence \(\mathbf{x}\) and previously generated tokens \((y_1,...,y_{i-1})\), i.e.,  
\[
P(y_i|\mathbf{x},y_1,...,y_{i-1})
\]

To maintain consistency with different model architectures, we adopt their respective predefined formatting conventions. For CodeLlama-7B/13B and Mistral, we use \texttt{[INST]} as the start token and \texttt{[/INST]} as the separator between the repaired code and its explanation. For StarChat-alpha, we follow its system-based prompting format, where the dialogue begins with \texttt{<|system|>\textbackslash n<|end|>\textbackslash n<|user|>} and is segmented using \texttt{<|end|>\textbackslash n<|assistant|>}. Each sequence is terminated with the model's default predefined end-of-sequence (EOS) token.

The computation of \(Loss_{1}\) uses a cross-entropy loss function to evaluate the discrepancy between the LLM's predicted probability distribution \(P(y_i|\mathbf{x}, y_1, ..., y_{i-1})\) for each token in the sequence and the actual distribution \(Q(y_i|\mathbf{x}, y_1, ..., y_{i-1})\), defined as:

\begin{equation}
\begin{split}
Loss_{1} = -\sum_{i} & Q(y_i|\mathbf{x},y_1,...,y_{i-1}) \\
            & \times \log P(y_i|\mathbf{x},y_1,...,y_{i-1})
\end{split}
\end{equation}

For \(Loss_{2}\), which assesses the LLM's capability to generate relevant explanatory guidance alongside repaired code, the loss calculation extends to the entire sequence of both code and guidance tokens, using a similar cross-entropy function:

\begin{equation}
\begin{split}
Loss_{2} =  -\sum_{i} & Q'(y_i|\mathbf{x},y_1,...,y_{i+n}) \\
           & \times \log P(y_i|\mathbf{x},y_1,...,y_{i+n})
\end{split}
\end{equation}

Here, $n$ represents the number of guidance tokens added to the sequence, aiming to enhance the LLM's ability to provide comprehensive expected outputs.

\subsection{Repair Inference}
\label{subsec:inference}
In the final step of our approach, we combine quantized LLM with QLoRA adapters to generate repaired codes during inference. The buggy code, represented by instruction $x$, is first transformed into a vector representation $\mathbf{x}=(x_1,...,x_n)$, where each $x_i$ corresponds to the embedding vector of the $i$-th token. This vector $\mathbf{x}$ is fed into fine-tuned LLMs equipped with quantization and QLoRA adapters to facilitate efficient and precise program repair generation. The computation within each linear layer of the quantized LLM is performed as follows:

\begin{equation}
Y = X \cdot \text{doubleDequant}(c_1, c_2, W) + X \cdot L_1 \cdot L_2
\end{equation}

Here, $\text{doubleDequant}(\cdot)$ functionally restores the quantized weight matrix $W$ to its complete precision, and $L_1, L_2$ are the QLoRA adapter matrices.

Through dequantization, we ensure that computations achieve the necessary precision for high-quality output, with each layer's output feeding into the subsequent layer until a final probability distribution over the vocabulary is achieved. This distribution, $p(y_i|\mathbf{x},y_1,...,y_{i-1})$, guides the generation of the next tokens $y_1,...,y_{i-1}$, based on both the input $\mathbf{x}$ and the previously generated tokens $y_1,...,y_{i-1}$. We employ sampling techniques and hyper-parameters to generate diverse and coherent program repairs, setting the temperature and $top_p$ (nucleus sampling) to $1.0$ and enabling token random sampling during generation. The resulting sequence $\hat{s}=(y_1,...,y_m)$ encompasses the LLM's response, from which we extract the first code segment delimited by triple backticks as the candidate repaired code $\hat{s}$. An example of this output format, including the extracted fixed code, is shown in Listing~\ref{lst:output_example}. This step is crucial because the LLM's output may encompass candidate-repaired code and supplementary natural language descriptions.

\begin{lstlisting}[language=mymarkdown,frame=tb,basicstyle=\footnotesize,label={lst:output_example}, caption= An example of output generated by CodeLlama-13B with \tool,numbers=none]
[INST] This is an incorrect code (add.cpp):
```c++
/*
Given a non-empty vector of integers lst. add the even elements that are at odd indices..
Examples:
    add({4, 2, 6, 7}) ==> 2
*/
#include<stdio.h>
#include<vector>
using namespace std;
int add(vector<int> lst){
    int sum=0;
    for (int i=0;i*2+1<lst.size();i++)
        if (lst[i*2+1]%2==1) sum+=lst[i*2+1];
    return sum;
}
```
You are a software engineer. Can you repair the incorrect code?
[/INST]
```c++
#include<stdio.h>
#include<vector>
using namespace std;
int add(vector<int> lst){
    int sum=0;
    for (int i=0;i*2+1<lst.size();i++){
        if (lst[i*2+1]%2==0) sum+=lst[i*2+1];
    }
    return sum;
}
```
Certainly. Let's go through the steps to fix the incorrect code:
1. The function is supposed to add all the even integers from the vector, but it only adds odd ones. We need to invert the condition to check for even numbers.
2. The function only checks for the elements in odd indices; it should check for all elements from index 0 to `lst.size()-1` (inclusive).
3. The function should return the sum of the even integers, but it returns `0` if no even integers are found. This is the behavior we want, but make sure that the function indeed returns `0` in this case.
By following these steps, we can correct the code to meet the desired functionality.
Remember to compile and test the code after making these changes to ensure that it works as expected.
\end{lstlisting}
\section{Experimental Setup}
\label{studydesign}
First, we outline the LLMs utilized in this study. Next, we detail the benchmark used for evaluation in our experiments. Following this, we explain the metrics employed to evaluate the repair capabilities of the fine-tuned LLMs. Then, we provide detailed hyper-parameters during fine-tuning. Lastly, we list the research questions we aim to address through this study.

\subsection{Model Selection}

To evaluate the generalizability of our approach, it is crucial to experiment with LLMs of varying architectures and sizes. Given the significant computational resources required for training and deploying large-scale LLMs, as highlighted by Chen~\etal~\cite{chen2022empowering}, we focused on code-targeted LLMs with parameter range of 7B to 16B. Table \ref{tab:models} presents our selected models, chosen based on their popularity~(as indicated by downloads from HuggingFace) and the diversity of their underlying architectures. These LLMs include CodeLlama-13B-instruct~\cite{roziere2023code}, CodeLlama-7B-instruct~\cite{roziere2023code}, StarChat-alpha~\cite{Tunstall2023starchat-alpha}, and Mistral-Instruct-7B~\cite{jiang2023mistral}, allowing us to comprehensively assess our approach's efficacy.

\begin{table}[H]
    \centering
    \caption{Selected Models}
    \resizebox{.8\textwidth}{!}{
        \begin{threeparttable}
            \begin{tabular}{|l|lrr|}
                \hline
                Model & Base Model & \# Params  & Downloads\tnote{*} \\
                \hline
                CodeLlama-13B-instruct & CodeLlama & 13B & \ 46.4k \\
                CodeLlama-7B-instruct & CodeLlama & \ 7B & \ 59.5k  \\
                StarChat-alpha & StarCoderBase & 16B & \ 24.9k \\
                Mistral-Instruct-7B-v0.1 & Mistral-7B-v0.1 & \ 7B  & 773.6k \\
                \hline
            \end{tabular}
            \begin{tablenotes}
                \small
                \item [*] ``Downloads'' count reflects the number of times LLMs were downloaded from HuggingFace before Feb. 2024.
            \end{tablenotes}
        \end{threeparttable}
    }
    \label{tab:models}
\end{table}

The selected models represent a variety of architectures and sizes: CodeLlama-13B-instruct and CodeLlama-7B-instruct, building on the Llama2 architecture~\cite{touvron2023llama}, offer infilling capabilities and optimized large-batch inference, demonstrating the adaptability of the CodeLlama~\cite{roziere2023code} foundation. StarChat-alpha, based on StarCoder~\cite{li2023starcoder}, introduces advanced pre-training techniques and benefits from expansive code datasets such as The Stack~\cite{khlaaf2022thestack}, illustrating a novel approach to leveraging data diversity for performance gains. Meanwhile, Mistral-Instruct-7B-v0.1, based on Mistral~\cite{jiang2023mistral}, emphasizes advancements in attention mechanisms, highlighting the potential for auto-regressive models in processing long sequences efficiently. In the following paragraphs, we denote CodeLlama-13B-instruct as CodeLlama-13B, CodeLlama-7B-instruct as CodeLlama-7B, and Mistral-Instruct-7B-v0.1 as Mistral-7B.

\subsection{Evaluation Benchmark}
\label{benchmark}
To rigorously assess the performance of our code-related framework, we aimed to establish robust program repair benchmarks in multiple programming languages. This is important because multiple languages are commonly used in practice, even within the same project~\cite{luo2025unlockingllmrepaircapabilities}. Therefore, evaluating cross-language generalizability is crucial. We focused on C++ and Java due to their popularity in the industry and automated program repair research.

We started with HumanEval-X~\cite{zheng2023codegeex}, a multilingual extension of the HumanEval benchmark~\cite{chen2021humaneval}. HumanEval-X expands each of the original 164 problems in HumanEval, initially in Python, to include equivalent problems in other languages, including C++ and Java. We chose HumanEval-X for two main reasons. First, it has a low risk of data leakage compared to datasets from before LLM training cutoffs (like Defects4J). Second, it affords a controlled study by providing the same problems in multiple languages. This allows us to attribute performance differences between languages to the languages themselves rather than differences in programs, problems, or data leakage.

To convert HumanEval-X into a suitable benchmark for evaluating program repair, we adapted the methodology from Jiang~\etal~\cite{jiang@impact}, which was used to create HumanEval-Java. We injected various bugs into the C++ and Java sections of HumanEval-X to form the EvalRepair-C++ and EvalRepair-Java benchmarks, respectively. This methodology ensures that the introduced bugs are equivalent across the two languages. For the bugs in EvalRepair-Java, we directly adopted the same bug patterns used in HumanEval-Java. As shown in Table \ref{tab:dataset_statistic}, although C++ and Java languages may significantly differ in implementing the same logic, the final distribution of single and multi-hunk bugs is almost consistent.

\begin{table}[h]
    \centering
    \caption{Statistics of EvalRepair-C++ and EvalRepair-Java}
    \begin{tabular}{|l|rr|}
        \hline
         & EvalRepair-C++ & EvalRepair-Java \\
        \hline
        bugs & 164 & 163 \\
        single-hunk bugs & 143 & 144\\
        multi-hunk bugs & 21 & 19\\
        average functions & 1.02 & 1.07 \\
        test cases & 96799 & 95677 \\
        \hline
    \end{tabular}
    \label{tab:dataset_statistic}
\end{table}

However, we recognized the potential for overfitting due to limited test cases. To address this, we enriched EvalRepair-C++ and EvalRepair-Java by combining both original and additional test cases from EvalPlus~\cite{liu2023your}, which we translated to C++ and Java. This expansion revealed that some original solutions in HumanEval-X failed to pass the new, more rigorous test cases. We corrected these issues in both languages to ensure the correctness of the augmented benchmarks. As a result, the average number of test cases per problem increased significantly to 590 and 587 for EvalRepair-C++ and EvalRepair-Java, respectively. This augmentation actively mitigates patch overfitting issues and provides a more accurate assessment of model performance.

\begin{table}[h]
    \centering
    \caption{Mitigation of Patch Overfitting}
    \begin{tabular}{|l|rr|}
        \hline
         & EvalRepair-C++ & EvalRepair-Java \\
        \hline
        \ding{172} \# Original Test Cases & 7 & 7 \\
        \ding{173} \# Augmented Test Cases & 590 & 587 \\
        \hline
        CodeLlama-13B Top-10 with \ding{172} & 67.7 & 73.6\\
        CodeLlama-13B Top-10 with \ding{173} & 58.5 & 69.9 \\
        \hline
    \end{tabular}
    \label{tab:dataset}
\end{table}

As illustrated in Table \ref{tab:dataset}, augmenting the test cases leads to a noticeable decline in the Top-10 of LLMs such as CodeLlama-13B, which experienced a reduction of 9.2\% in EvalRepair-C++ and 3.7\% in EvalRepair-Java. Introducing a more comprehensive set of test cases highlights the importance of rigorous evaluation in developing LLMs. It sets a new standard for assessing their performance in program repair tasks. Test cases are used only for evaluation - our model generates patches without leveraging any test information. These benchmarks, EvalRepair-C++ and EvalRepair-Java, will be made publicly accessible via an API, ensuring that the research community can benefit from these resources for future explorations and improvements in the field without data leakage problems.

\subsection{Evaluation Metrics}
\label{metrics}
To accurately evaluate the effectiveness of LLMs in program repair, this study employs three primary metrics: Top-1, Top-5, and Top-10. The ``Top-k'' metric is the scenario where, among the top k candidate patches produced by the LLMs, the code is considered successfully repaired if any candidates pass all test cases in the benchmark. This metric selection is grounded in the observation by Kochhar~\etal\cite{kochhar2016practitioners} that most developers tend to abandon automated debugging tools if they fail to identify the actual bugs within the first five attempts. Furthermore, Noller~\etal~\cite{noller2022trust} found that developers are unlikely to consider more than the top-10 ranked patches when seeking solutions. Reflecting on these insights and aligning with the findings from prior program repair studies~\cite{huq2022review4repair,fu2022vulrepair,noller2022trust,wong2021varfix}, our selection of the Top-1, Top-5, and Top-10 metrics is not only justified but also crucial for ensuring our evaluation mirrors real-world developer scenarios and expectations.

\subsection{Fine-tuning Configurations}
\label{hyper-parameters}
\begin{figure}[htbp]
   \centering
   \begin{subfigure}[b]{0.45\textwidth}
       \centering
       \includegraphics[width=\textwidth]{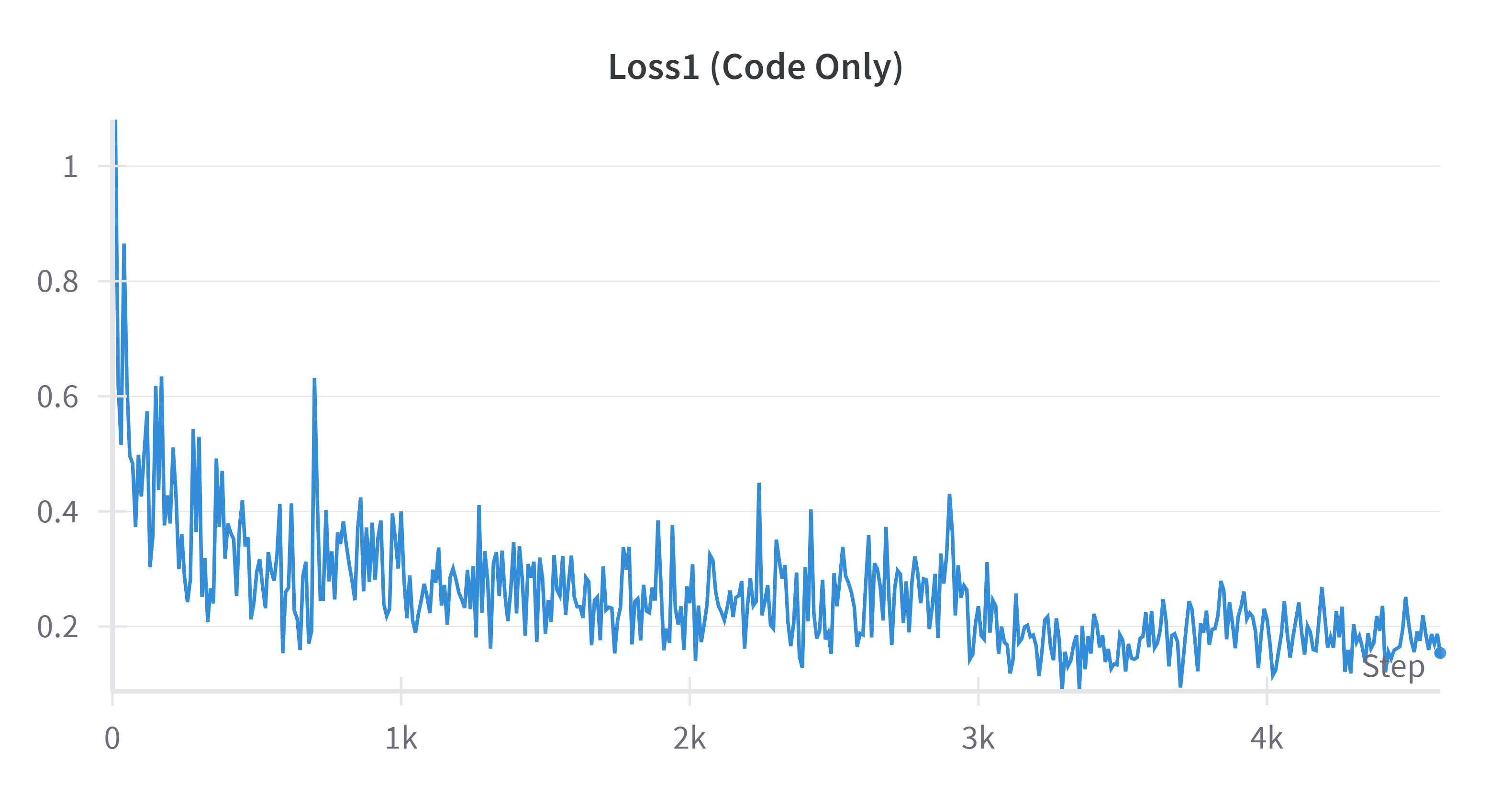}
       \caption{\(Loss_1\) values across iterations.}
       \label{fig:loss1}
   \end{subfigure}
   \hfill
   \begin{subfigure}[b]{0.45\textwidth}
       \centering
       \includegraphics[width=\textwidth]{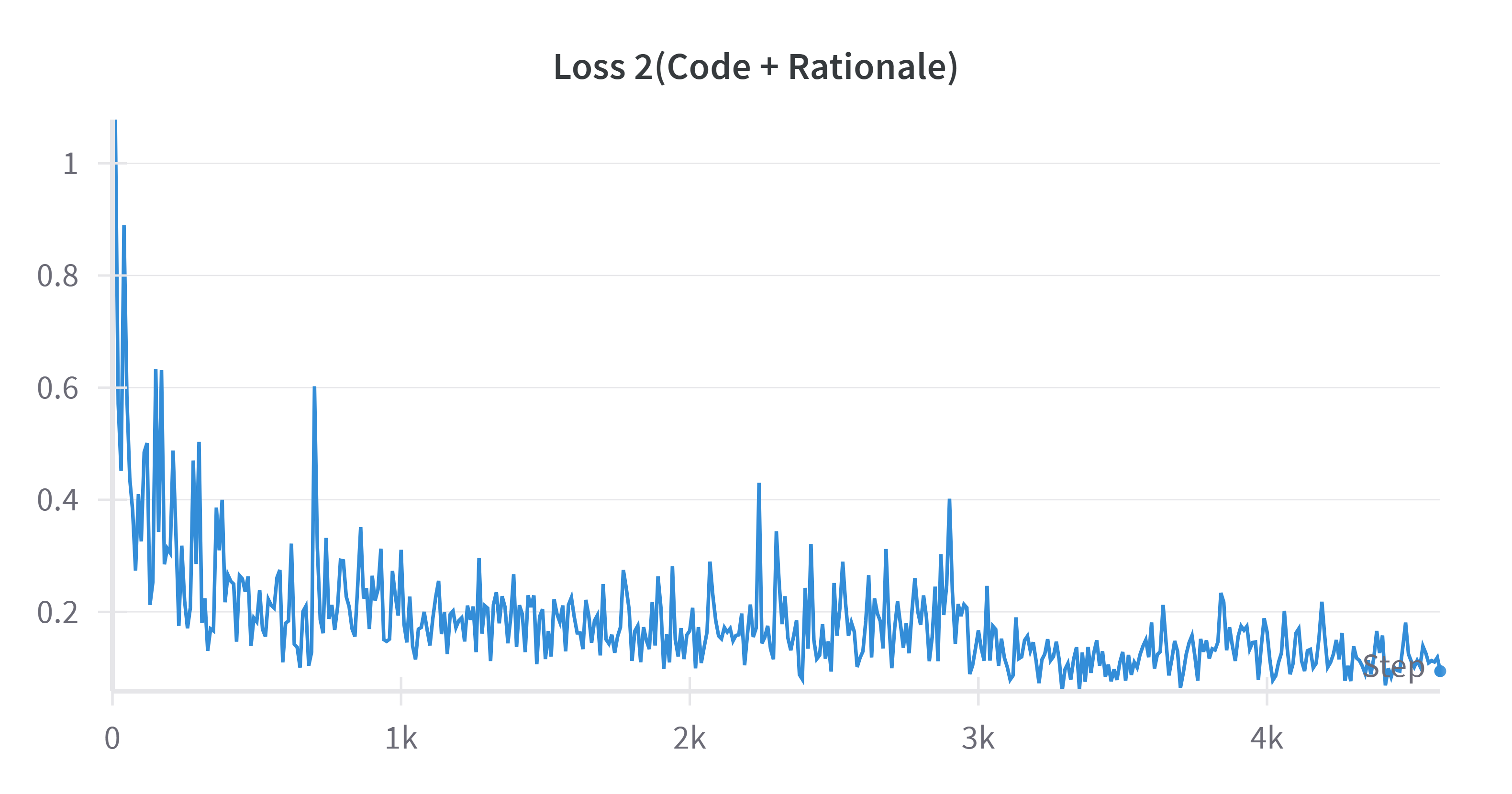}
       \caption{\(Loss_2\) values across iterations.}
       \label{fig:loss2}
   \end{subfigure}
   \caption{The comparison of \(Loss_1\) and \(Loss_2\) across iterations.}
   \label{fig:loss_comparison}
\end{figure}

To balance multi-objective optimization, we adopt a weighted sum formulation in Equation~\ref{eq:loss} and set \(\alpha=0.5\), following established practices in multi-objective learning~\cite{kokkinos2017ubernet,hsieh2023distilling}. Since \(Loss_1\) and \(Loss_2\) share a common prefix and exhibit comparable magnitudes during training (Figure~\ref{fig:loss_comparison}), this equal weighting configuration avoids optimization-induced variance. It ensures a more transparent evaluation of multi-objective fine-tuning.

For parameter-efficient fine-tuning, all approaches in our experiments, including \tool, Fine-tune-CoT, and standard fine-tuning (\stdft), leverage LoRA for efficient adaptation. Additionally, we incorporate NEFTune to prevent overfitting and improve the generalization of fine-tuned models. \tool optimizes two objectives: (1) generating repaired code and (2) generating repaired code along with natural language rationales. \stdft optimizes solely for the first objective, while Fine-tune-CoT optimizes for the second. This ensures that any performance differences are caused by differences in the fine-tuning process rather than differences in optimization techniques. To further maintain experimental consistency and efficiency, we apply the following hyper-parameter settings across all fine-tuning methods:

\begin{itemize}
   \item \textbf{LoRA rank \(r=32\) and scaling factor \(LoRA\_alpha=16\)}: LoRA~\cite{hu2021lora} is employed in all fine-tuning methods (\tool, \stdft, and Fine-tune-CoT) to ensure fair comparisons. It introduces low-rank updates to efficiently fine-tune large models, where \(r\) controls adaptation capacity, and \(LoRA\_alpha\) determines their influence on pre-trained weights. Following~\cite{wang2024lora,raschka2024practical}, we set \(r=32\) and \(LoRA\_alpha=0.5r\), balancing efficiency and performance. QLoRA further reduces training overhead by applying LoRA to a quantized model~\cite{dettmers2023qlora}. In our experiments, it decreases the number of trainable parameters from 6,672,143,360 to 125,173,760 (1.88\%), significantly lowering memory consumption.

   \item \textbf{NEFTune noise scale \(NEFTune\_alpha=5.0\)}: NEFTune introduces stochastic noise into embeddings, scaled by \(NEFTune\_alpha / \sqrt{L \cdot d}\), where \(L\) is the sequence length and \(d\) is the embedding dimension. This improves generalization, with our setting of \(NEFTune\_alpha=5.0\) yielding the best accuracy improvement on most repair benchmarks~\cite{jain2023neftune}.

   \item \textbf{Context length 2048}: This setting captures sufficient context for repair tasks while remaining within CodeLlama’s 4096 token limit~\cite{roziere2023code}.
\end{itemize}

\subsection{Research Questions}
\textbf{RQ-1: How effective is fine-tuning with two objectives for program repair?} We investigate the performance of \tool's multi-objective fine-tuning in contrast to standard, single-objective fine-tuning on the CodeLlama-13B, which has been evidenced as achieving state-of-the-art performance for program repair tasks~\cite{zirak2022improving,huang2023empirical}. This comparative analysis is conducted using the EvalRepair-C++ and EvalRepair-Java benchmarks to assess not only the effectiveness of \tool in improving program repair but also its ability to generalize across different programming languages.

\textbf{RQ-2: How does model size or type impact repair performance of \tool?} We examine \tool's performance on LLMs with distinct sizes and architectures, including CodeLlama-13B, CodeLlama-7B, StarChat-alpha-16B, and Mistral-7B, on EvalRepair-C++ and EvalRepair-Java benchmarks. This study aims to validate \tool's generalization capability by comparing its fine-tuning effects against standard approaches and baseline performance across varying LLMs.

\textbf{RQ-3: How does \tool compare against \tool with human guidance and state-of-the-art fine-tuning methods?} Through an ablation study, we explore the influence of the source of guidance~(LLM-generated vs. human-generated) on \tool's effectiveness. Additionally, we compare \tool’s performance to two advanced fine-tuning methodologies, Fine-tune-CoT~\cite{ho2022large} and RepairLLaMA~\cite{silva2023repairLlama}, across various LLMs.

\textbf{RQ-4: How well does \tool perform on repository-level bugs?} We extend our evaluation to subsets of repository-level benchmarks, including Defects4J~\cite{just2014defects4j} and SWE-Bench~\cite{jimenez2023swe}, to assess \tool's effectiveness in repairing real-world bugs. By comparing \tool's performance with standard fine-tuning on these benchmarks, we aim to evaluate its scalability and ability to handle complex real-world repair scenarios.

\section{Experiments \& Results}
\label{experiments}
\subsection{Effectiveness of Multi-objective Fine-tuning for Program Repair}

\noindent
\textbf{[Objective:]} This study assesses \tool's impact on fine-tuning LLMs for program repair, comparing its multi-objective approach against standard single-objective fine-tuning and baseline LLMs without fine-tuning. Our investigation centers around two sub-questions:

\begin{itemize}
    \item {\bf RQ-1.1} { How does fine-tuning LLMs with \tool compare to both standard fine-tuning and the baseline LLM in terms of repair performance? }
    \item {\bf RQ-1.2} { Does \tool exhibit cross-language generalization in program repair tasks compared to standard fine-tuning and baseline LLM? }
\end{itemize}

\noindent
\textbf{[Experimental Design for RQ-1.1]:} We fine-tune CodeLlama-13B using both \tool and the standard fine-tuning approach. Here, the baseline represents CodeLlama-13B without any fine-tuning, serving as our control for evaluating the impact of fine-tuning. Standard fine-tuning refers to fine-tuning CodeLlama-13B to generate repaired code without other information, denoted as \stdft. In contrast, \tool involves multi-objective fine-tuning, aiming to enhance LLM's repair capabilities through additional natural language guidance. The comparative analysis is based on Top-1, Top-5 and Top-10 metrics on the benchmark EvalRepair-C++, detailed in Section \ref{benchmark}. As shown in Listing~\ref{lst:gpt_prompt_cpp}, we use a simple prompt template for evaluation.

\begin{lstlisting}[language=mymarkdown,frame=tb,basicstyle=\footnotesize, label={lst:gpt_prompt_cpp}, caption=Prompt template for evaluating models on EvalRepair-C++ benchmark,numbers=none]
This is an incorrect code(Filename.cpp):
```cpp
Here is the cpp code
```
You are a software engineer. Can you repair the incorrect code?
\end{lstlisting}

\noindent
\textbf{[Experimental Results for RQ-1.1]:} Table \ref{tab:rq1.1} shows \tool's significant repair performance enhancement on EvalRepair-C++ over both the baseline and \stdft. Against the baseline, \tool elevates Top-5 by 20.7 percentage points~(a 50.6\% relative increase), and Top-10 by 11.0 percentage points~(an 18.8\% relative increase). Compared to \stdft, \tool maintains its superiority with increments of 12.2 percentage points~(a 24.7\% relative increase) in Top-5, and 5.5 percentage points~(an 8.6\% relative increase) in Top-10. These substantial improvements, particularly in Top-5 and Top-10, more than double the gains of \stdft, showcasing \tool's profound impact. The success of the \tool approach underscores the advantage of multi-objective fine-tuning in fostering a more nuanced understanding and application of repair logic than what is achieved through standard fine-tuning.

\begin{table}[h]
    \centering
    \caption{Fine-tune CodeLlama-13B with \stdft and \tool vs GPT-4 on EvalRepair-C++.}
    \begin{tabular}{|l|lll|}
        \hline
        Model & Top-1 & Top-5 & Top-10 \\
        \hline
        GPT-4 & 67.1 & 97.6 & 98.2 \\
        \hline
        CodeLlama-13B & 12.2 & 40.9 & 58.5 \\
        CodeLlama-13B-\stdft & 22.0~(+ 9.8) & 49.4~(+ 8.5) & 64.0~(+ 5.5)\\
        CodeLlama-13B-\tool & \textbf{28.7~(+16.5)} & \textbf{61.6~(+20.7)} & \textbf{69.5~(+11.0)}  \\
        \hline
    \end{tabular}
    \label{tab:rq1.1}
\end{table}

\noindent
\textbf{[In-depth Analysis for RQ-1.1]:} We analyze two repair cases that highlight \tool's capabilities and limitations.

\noindent
\textbf{Successful Repair (derivative.cpp).} We examine a representative sample from EvalRepair-C++, ``derivative.cpp'', where \tool successfully generated a correct patch while both the baseline and \stdft failed. As shown in Listing~\ref{lst:buggy-derivative}, the buggy implementation simply copies the input coefficients without performing derivative calculations.

\begin{lstlisting}[language=C++,frame=tb,basicstyle=\footnotesize,label={lst:buggy-derivative}, caption=Buggy implementation of derivative function,numbers=none]
vector<float> derivative(vector<float> xs){
    vector<float> out={};
    for (int i=0;i<xs.size();i++)
        out.push_back(xs[i]);
    return out;
}
\end{lstlisting}

The key challenge in this bug is applying polynomial derivative rules correctly. The derivative of a polynomial term $ax^n$ is $nax^{n-1}$, meaning each coefficient must be multiplied by its corresponding power. The baseline model (Listing \ref{lst:baseline-patch}) incorrectly starts indexing from 0 instead of 1, leading to incorrect coefficient calculations.

\begin{lstlisting}[language=C++,frame=tb,basicstyle=\footnotesize,label={lst:baseline-patch}, caption=Baseline's incorrect patch for derivative,numbers=none]
vector<float> derivative(vector<float> xs){
    for (int i=0;i<xs.size()-1;i++) {
        out.push_back(xs[i]*i);
    }
    return out;
}
\end{lstlisting}

\stdft introduced unnecessary complexity and incorrect operations, as shown in Listing \ref{lst:stdft-patch}. It modifies the input vector instead of creating a new output, unnecessarily removes zero coefficients, and includes an incorrect division operation.

\begin{lstlisting}[language=C++,frame=tb,basicstyle=\footnotesize,label={lst:stdft-patch}, caption=\stdft's incorrect patch for derivative,numbers=none]
vector<float> derivative(vector<float> xs){
    for (int i=0;i<xs.size();i++) {
        if (xs[i]==0) {
            xs.erase(xs.begin() + i);
            i--;
        } else {
            xs[i]=xs[i]*i;
            xs[i]=xs[i]/i-1;
        }
    }
    return xs;
}
\end{lstlisting}

In contrast, \tool identified the issue and generated the expected patch~(Listing ~\ref{lst:morepair-patch}), iterating from index 1 and correctly applying the derivative formula.

\begin{lstlisting}[language=C++,frame=tb,basicstyle=\footnotesize,label={lst:morepair-patch}, caption=\tool's correct patch for derivative,numbers=none]
vector<float> derivative(vector<float> xs){
    vector<float> out;
    int i;
    for (i=1;i<xs.size();i++){
        out.push_back(xs[i]*i);
    }
    return out;
}
\end{lstlisting}

Furthermore, \tool provided an accurate explanation:

\begin{quote}
\textit{The code was missing multiplication by the current position to get the term's coefficient when computing the derivative. The loop should iterate from `i=1` to `i<xs.size()` and multiply `xs[i]` by `i` to get the coefficient of the term corresponding to derivative powers `i-1`.}
\end{quote}

This demonstrates \tool's deep understanding of both the mathematical concept and its implementation, directly contributing to the generation of the correct patch.

\noindent
\textbf{Failure Repair (all\_prefixes.cpp).} We present a representative case where \tool failed to generate a correct patch. The ``all\_prefixes.cpp'' function is intended to generate all prefixes of a given string. However, the incorrect implementation~(Listing \ref{lst:buggy-allprefix}) contains an ``index out-of-bounds error'', where accessing \texttt{str[i + 1]} in the loop exceeds the string length when \texttt{i = str.length() - 1}, leading to undefined behavior.

\begin{lstlisting}[language=C++,frame=tb,basicstyle=\footnotesize,label={lst:buggy-allprefix}, caption=Buggy implementation of all\_prefixes function,numbers=none]
vector<string> all_prefixes(string str){
    vector<string> out;
    string current="";
    for (int i=0;i<str.length();i++)
    {
        current=current+str[i+1];
        out.push_back(current);
    }
    return out;
}
\end{lstlisting}

\tool produced an incorrect patch~(Listing \ref{lst:morepair-patch1}), which does not correctly preserve the incremental construction of string prefixes and introduces an off-by-one logic error.

\begin{lstlisting}[language=C++,frame=tb,basicstyle=\footnotesize,label={lst:morepair-patch1}, caption=\tool's incorrect patch for all\_prefixes,numbers=none]
vector<string> all_prefixes(string str){
    vector<string> out;
    for(int i=0;i<=str.length();i++){
        out.push_back(str.substr(0,i));
    }
    return out;
}
\end{lstlisting}

The generated patch attempts to use \texttt{substr(0, i)} to construct prefixes but introduces another off-by-one error, failing to match the expected behavior. While \tool's first patch was incorrect, it did successfully fix this bug in its Top-10 results. Moreover, \tool successfully produced an insightful and accurate natural language rationale:

\begin{quote}
\textit{1. **Identify the problem**: The original code is iterating over the characters of the string, but it's not properly handling the index out of bounds.}

\textit{2. **Analyze the loop condition**: The loop is iterating from `0` to `str.length()`, but inside the loop, `current` is being set to `current + str[i+1]`. This will go out of bounds if `i` is equal to `str.length() - 1`.}

\textit{3. **Fix the loop condition**: Change the loop condition to iterate from `0` to `str.length() - 1`, as the maximum index is one less than the length of the string.}

\textit{4. **Correct the loop body**: Adjust the loop body to account for the corrected loop condition. You will need to modify `current=current+str[i+1]` to `current=current+str[i]` and ensure that the final prefix is also included in the result.}

\textit{5. **Test the code**: After making these changes, test the code with various input strings to ensure that it works as expected.}

\textit{By following these steps, you should be able to correct the code and achieve the desired functionality.}
\end{quote}

This case demonstrates that even when \tool does not generate a correct patch, its precise and structured NL rationale offers meaningful debugging guidance. Developers can leverage this explanation to identify and correct the error efficiently~\cite{tang2024codeagent}.

\noindent
\textbf{Manual Statistics.} To evaluate the relationship between \tool's repair accuracy and its ability to generate meaningful natural language (NL) explanations, we manually analyzed its Top-1 repair results on EvalRepair-C++. This analysis quantifies cases where \tool provides useful guidance even when the generated patch is incorrect.

Table~\ref{tab:tool_manual_analysis} summarizes the findings. Among 164 repair attempts, \tool produced correct NL explanations in 45 cases (27.4\%), of which 36 (80.0\%) had correct patches, while 9 (20.0\%) were incorrect. In these 9 cases, \tool correctly identified the repair location and described the issue accurately, but the code itself failed due to minor implementation details, causing it to fail some test cases. However, all instances where \tool generated correct NL but incorrect fixed code in Top-1 had at least one correct fix in the Top-10 results.

Of the 11 cases with correct fixes but incorrect guidance, 3 had empty guidance, 6 repeated the fixed code, 1 provided guidance in Korean, and 1 generated redundant test cases. This highlights that the guidance is either correct or ineffective when the fix is correct. Accurate guidance is essential for understanding and applying the fix, as it ensures developers can adequately interpret the solution, even if the code itself is correct. Among the 117 incorrect fixes of \tool, 9 (7.7\%) were paired with correct guidance, which can still be helpful to developers, as it provides valid repair suggestions even when the fix itself is incorrect. In conclusion, by selecting a fix that passes all test cases and has valid guidance, we can reliably obtain a result where both the guidance and the fixed code are correct~\cite{tian2022change}.

\begin{table}[h]
    \centering
    \caption{Manual analysis of \tool's Top-1 results on EvalRepair-C++.}
    \begin{tabular}{|l|ll|l|}
        \hline
        & Correct Fix & Wrong Fix & Total \\
        \hline
        Correct NL & 36 & 9 & 45 \\
        Wrong NL & 11 & 108 & 119 \\
        \hline
        Total & 47 & 117 & 164 \\
        \hline
    \end{tabular}
    \label{tab:tool_manual_analysis}
\end{table}

To further analyze \tool's performance across different bug types, we categorize EvalRepair-C++ into three categories.

\begin{enumerate}
    \item \textbf{Condition errors} (\textit{n}=58): Incorrect loop bounds or conditional checks.
    \item \textbf{Expression errors} (\textit{n}=65): Incorrect assignments or arithmetic expressions.
    \item \textbf{Structural errors} (\textit{n}=41): Errors that span multiple logic blocks or entire if/for structures, requiring coordinated modifications across different parts of the program to fix logical bugs.
\end{enumerate}

\begin{table}[h]
    \centering
    \caption{Top-1 accuracy of CodeLlama-13B on Fixing different types of bugs in EvalRepair-C++.}
    \begin{tabular}{|l|l|lll|}
        \hline
        Bug Type & \#Count & CodeLlama-13B & \stdft & \tool  \\
        \hline
        Condition Errors & 58 & 12.1 & 17.2 & 25.9 \\
        Expression Errors & 65 & 15.4 & 23.1 & 20.0 \\
        Structural Errors & 41 & 7.3 & 26.8 & 46.3 \\
        \hline
    \end{tabular}
    \label{tab:bug_complexity_analysis1}
\end{table}

Table \ref{tab:bug_complexity_analysis1} presents the Top-1 accuracy of CodeLlama-13B, \stdft, and \tool across three types of bugs in EvalRepair-C++. Notably, for Structural Errors, both \stdft and \tool show significant improvements over the baseline, with \tool outperforming \stdft by 72.8\%. This highlights \tool’s superior ability to handle complex logical bugs, similar to other reasoning models~\cite{guo2025deepseek}. Additionally, \tool achieves the highest minimum accuracy across all bug types, indicating that multi-objective learning and LLM-generated guidance enhance its generalization performance.

\noindent
\textbf{[Experimental Design for RQ-1.2]:} To probe \tool's and \stdft's capacity for cross-language generalization in program repair, we fine-tuned CodeLlama-13B with each method using the C++ training dataset \dataset and evaluated them on the Java repair benchmark EvalRepair-Java across Top-1, Top-5, and Top-10 metrics, offering insights into how these approaches adapt to a language different from the training dataset. We employ a simple prompt template for evaluation, as illustrated in Listing~\ref{lst:gpt_prompt_java}.

\noindent
\textbf{[Experimental Results for RQ-1.2]:} The repair performance presented in Table \ref{tab:rq1.2} for the EvalRepair-Java benchmark details how both \stdft and \tool extend their capabilities into a cross-language scenario. \stdft enhances the Top-10 by 9.7\% over the baseline~(CodeLlama-13B), while \tool further improves upon this, exhibiting an additional 1.6\% increase in Top-10 over \stdft. These enhancements validate the cross-language generalization capability of both fine-tuning approaches, with \tool showcasing superior performance in adapting to Java, which is a shift from the training dataset's programming language. Notably, \tool achieves a 77.9\% Top-10, marking an 11.4\% increase over the baseline. This significant improvement underscores \tool's effectiveness in cross-language repair scenarios.

\begin{table}[h]
    \centering
    \caption{Fine-tune CodeLlama-13B with \stdft and \tool vs GPT-4 on EvalRepair-Java.}
    \begin{tabular}{|l|lll|}
        \hline
        Model &Top-1 & Top-5 & Top-10 \\
        \hline
        GPT-4 & 72.3 & 85.9 & 89.0 \\
        \hline
        CodeLlama-13B & 23.3 & 54.0 & 69.9 \\
        CodeLlama-13B-\stdft & 33.7~(+10.4) & 62.0~(+ 8.0) & 76.7~(+ 6.8) \\
        CodeLlama-13B-\tool & \textbf{35.0~(+11.7)} &\textbf{69.3~(+15.3)} & \textbf{77.9~(+ 8.0)} \\
        \hline
    \end{tabular}
    \label{tab:rq1.2}
\end{table}

\begin{lstlisting}[language=mymarkdown,frame=tb,basicstyle=\footnotesize, label={lst:gpt_prompt_java}, caption=Prompt template for evaluating models on EvalRepair-Java benchmark,numbers=none]
This is an incorrect code(Filename.java):
```java
Here is the java code
```
You are a software engineer. Can you repair the incorrect code?
\end{lstlisting}

Furthermore, we include GPT-4, state-of-the-art closed-source LLM, as a benchmark for upper limits of repair performance, as illustrated in Tables \ref{tab:rq1.1} and \ref{tab:rq1.2}. The result shows that \tool narrows the performance gap between CodeLlama-13B and GPT-4. To ensure a fair comparison, we adopt the same prompting format~(Listing \ref{lst:gpt_prompt_cpp} and \ref{lst:gpt_prompt_java}) and inference settings for GPT-4 as for the models evaluated in RQ-1. Specifically, we maintain consistency with the inference configuration detailed in Section \ref{subsec:inference}, setting temperature = 1.0, max\_token = 1024, top\_p = 1.0, and enabling token random sampling during generation.

\noindent
\textbf{[In-depth Analysis for RQ-1.2]:} To evaluate \tool's cross-language repair capabilities and limitations, we analyze a representative case from EvalRepair-Java. We analyze a case where \tool produced a correct patch with an insightful explanation, whereas \stdft generated an overly complex and incorrect solution. The buggy implementation~(Listing~\ref{lst:buggy-triples}) fails to enforce the constraint \texttt{i < j < k} in the nested loops.

\begin{lstlisting}[language=Java,frame=tb,basicstyle=\footnotesize,label={lst:buggy-triples}, caption=Buggy implementation of get\_max\_triples function,numbers=none]
public static int get_max_triples(int n) {
    int[] A = new int[n];
    for (int i = 1; i <= n; i += 1)
        A[i - 1] = i * i - i + 1;

    int result = 0;
    for (int i = 0; i < n-2; i += 1) {
        for (int j = 0; j < n-1; j += 1) {
            for (int k = 0; k < n; k += 1) {
                if ((A[i] + A[j] + A[k]) % 3 == 0) result += 1;
            }
        }
    }
    return result;
}
\end{lstlisting}

\stdft expanded the original 15-line implementation into 57 lines, introducing unnecessary complexity (Listing~\ref{lst:stdft-triples}), and did not generate an explanation.

\begin{lstlisting}[language=Java,frame=tb,basicstyle=\footnotesize,label={lst:stdft-triples}, caption=\stdft's over-engineered patch for get\_max\_triples,numbers=none]
public class GET_MAX_TRIPLES {
    // Additional 23 unnecessary lines
    public int get_max_triples(long n){               // Changed signature to use long
        int result = 0;
        for(long i = 1; i <= n-2; ++i){              // Direct computation without array
            for(long j = i+1; j <= n-1; ++j){        // Modified upper bound
                for(long k = j+1; k <= n; ++k){      // Modified upper bound
                    if(get_int(get_add(get_add(i*i-i,j*j-j), k*k-k), 3) == 0){
                        ++result;
                    }
                }
            }
        }
        return result;
    }
    // Additional 19 unnecessary lines
}
\end{lstlisting}

In contrast, \tool fixed the issue with only two modifications (Listing~\ref{lst:morepair-triples}).

\begin{lstlisting}[language=Java,frame=tb,basicstyle=\footnotesize,label={lst:morepair-triples}, caption=\tool's minimal correct patch for get\_max\_triples,numbers=none]
public static int get_max_triples(int n) {
    int[] A = new int[n];
    for (int i = 1; i <= n; i += 1)
        A[i - 1] = i * i - i + 1;
    int result = 0;
    for (int i = 0; i < n-2; i += 1) {
-       for (int j = 0; j < n-1; j += 1) {
+       for (int j = i+1; j < n-1; j += 1) {         // Fixed: ensure j > i
-           for (int k = 0; k < n; k += 1) {
+           for (int k = j+1; k < n; k += 1) {       // Fixed: ensure k > j
                if ((A[i] + A[j] + A[k]) % 3 == 0) result += 1;
            }
        }
    }
    return result;
}
\end{lstlisting}

\tool also provided a clear explanation:

\begin{quote}
\textit{After fixing the code, it should work correctly. The main change was to adjust the loops to ensure that `i < j < k` by starting the second loop from `i+1` and the third loop from `j+1`.}

\textit{The reason for this change is that the condition `i < j < k` only holds true if `i` is strictly less than `j` and `j` is strictly less than `k`. This means that once we set `j` to `i+1`, we are ensuring that both `i` and `j` are distinct. Then, when we set `k` to `j+1`, we are ensuring that `j` is strictly less than `k`.}
\end{quote}

This case highlights \tool's ability to generate minimal, targeted fixes while preserving the original structure. Despite being trained on C++ data, \tool correctly adapted to Java-specific syntax, demonstrating more substantial cross-language generalization than \stdft. This result, combined with \tool's higher Top-1, Top-5, and Top-10 results in Table~\ref{tab:rq1.2}, further reinforces its superior cross-language generalization over \stdft.

\find{{\bf [RQ-1] Findings:} (1) Fine-tuning with \tool outperforms CodeLlama-13B baseline significantly in repair performance. The improvements in the Top-10 for EvalRepair-C++ and EvalRepair-Java are 18.8\% and 11.4\%, respectively, showcasing superior repair capabilities. (2) Against \stdft, \tool shows repair performance gains with increases in Top-5 of 24.7\% for EvalRepair-C++ and 11.8\% for EvalRepair-Java, indicating generalization across programming languages. (3) Even when \tool fails to generate a correct patch, it can provide accurate guidance in 7.7\% of these cases, aiding developers in debugging. {\bf Insights:} (1) Our approach \tool highlights multi-objective learning's impact on automated program repair, proving its ability to enhance repair tasks. (2) \tool aids debugging by providing correct explanations even when patch generation fails.}
\subsection{Impact of Size or Type for Fine-tuning LLMs on Code Repair Performance}

\noindent
\textbf{[Objective]:} To investigate RQ-2, we assess the impact of fine-tuning with \tool on LLMs of varying sizes and architectures in terms of their code repair capabilities.

\noindent
\textbf{[Experimental Design]:} To examine the generalization of the \tool approach across LLMs with different sizes and architectures, we selected CodeLlama-7B, StarChat-alpha (which has 16B parameters), and Mistral-7B as our base LLMs. These LLMs represent a diverse range of architectures, and CodeLlama-7B differs in size from the CodeLlama-13B assessed in RQ-1. We fine-tune these LLMs using either standard fine-tuning~(\stdft) or \tool, then evaluate their performance on two benchmarks: EvalRepair-C++ and EvalRepair-Java.

\begin{table}[h]
    \centering
    \caption{Impact of model sizes or architectures on the effectiveness of fine-tuning on EvalRepair-C++.}
    \begin{threeparttable}
        \begin{tabular}{|l|lll|}
            \hline
            Model & Top-1 & Top-5 & Top-10 \\
            \hline
            CodeLlama-13B & 12.2 & 40.9 & 58.5 \\
            CodeLlama-13B-\stdft & 22.0~(+ 9.8) & 49.4~(+ 8.5) & 64.0~(+ 5.5)\\
            CodeLlama-13B-\tool & \textbf{28.7~(+16.5)} & \textbf{61.6~(+20.7)} & \textbf{69.5~(+11.0)}  \\
            \hline
            CodeLlama-7B & 15.2 & 46.3 & 59.1  \\
            CodeLlama-7B-\stdft & 19.5~(+ 4.3) & 50.0~(+ 3.7) & 61.6~(+ 2.5)  \\
            CodeLlama-7B-\tool & \textbf{24.4~(+ 9.2)} & \textbf{56.7~(+10.4)} & \textbf{62.8~(+ 3.7)} \\
            \hline
            StarChat-alpha & 18.3 & 50.0 & 62.2  \\
            StarChat-\stdft & 16.5~(- 1.8) & 43.3~(- 6.7) & 58.5~(- 3.7)\\
            StarChat-\tool & \textbf{23.8~(+ 5.5)} & \textbf{52.4~(+ 2.4)} & \textbf{65.9~(+ 3.7)} \\
            \hline
            Mistral-7B & 14.6 & 32.3 & 47.0 \\
            Mistral-7B-\stdft & 13.4~(- 1.2) & 39.0~(+ 6.7) & 46.3~(- 0.7) \\
            Mistral-7B-\tool & \textbf{16.5~(+ 1.9)} & \textbf{40.2~(+ 7.9)} & \textbf{50.0~(+ 3.0)} \\
            \hline
        \end{tabular}
        \begin{tablenotes}
            \small
            \item [†] Values in parentheses indicate the change relative to the corresponding baseline.
        \end{tablenotes}
    \end{threeparttable}
    \label{tab:rq2-1}
\end{table}

\noindent
\textbf{[Experimental Results]:} Table \ref{tab:rq2-1} outlines the Top-1, Top-5, and Top-10 repair performance metrics for baseline, \stdft, and \tool across four LLMs on EvalRepair-C++. Notably, \stdft doesn't consistently improve repair metrics, failing to surpass the repair performance of baseline on several base LLMs, such as StarChat-alpha. Conversely, \tool consistently enhances performance across all metrics and LLMs, with a maximum 18.8\% Top-10 improvement over baseline and a maximum 12.6\% Top-10 improvement over \stdft evaluated on EvalRepair-C++. This suggests superior generalizability of multi-objective learning across different LLMs for code repair.

\begin{table}[h]
    \centering
    \caption{Impact of model sizes or architectures on the effectiveness of fine-tuning on EvalRepair-Java.}
    \begin{threeparttable}
        \begin{tabular}{|l|lll|}
            \hline
            Model & Top-1 & Top-5 & Top-10 \\
            \hline
            CodeLlama-13B & 23.3 &  54.0 & 69.9  \\
            CodeLlama-13B-\stdft & 33.7~(+10.4) & 62.0~(+ 8.0) & 76.7~(+ 6.8) \\
            CodeLlama-13B-\tool & \textbf{35.0~(+11.7)} & \textbf{69.3~(+15.3)} & \textbf{77.9~(+ 8.0)} \\
            \hline
            CodeLlama-7B & 22.1& 49.7 & 62.0  \\
            CodeLlama-7B-\stdft & 20.2~(- 1.9)& 49.1~(- 0.6) & 60.7~(- 1.3) \\
            CodeLlama-7B-\tool & \textbf{22.7~(+ 0.6)} & \textbf{59.5~(+ 9.8)} & \textbf{67.5~(+ 5.5)}  \\
            \hline
            StarChat-alpha  &15.3 &  43.6 & 60.7   \\
            StarChat-\stdft &17.8~(+ 2.5) & 47.9~(+ 4.3) & 56.4~(- 4.3) \\
            StarChat-\tool & \textbf{27.6~(+12.3)} & \textbf{56.4~(+12.8)} & \textbf{66.3~(+ 5.6)} \\
            \hline
            Mistral-7B & 14.1 & 33.7 & 52.1 \\
            Mistral-7B-\stdft & 18.4~(+ 4.3) &  42.3~(+ 8.6) & 54.6~(+ 2.5) \\
            Mistral-7B-\tool & \textbf{19.0~(+ 4.9)} & \textbf{45.4~(+11.7)} & \textbf{58.3~(+ 6.2)} \\
            \hline
        \end{tabular}
        \begin{tablenotes}
            \small
            \item [†] Values in parentheses indicate the change relative to the corresponding baseline.
        \end{tablenotes}
    \end{threeparttable}
    \label{tab:rq2-2}
\end{table}

Table \ref{tab:rq2-2} presents the Top-1, Top-5, and Top-10 metrics for baseline, \stdft, and \tool on the EvalRepair-Java benchmark across four LLMs. Unlike the results from EvalRepair-C++ in Table \ref{tab:rq2-1}, CodeLlama-7B-\stdft under-performs on EvalRepair-Java, revealing \stdft's inconsistent cross-language generalization. Similarly, StarChat-\stdft's decline mirrors its performance on EvalRepair-C++, indicating \stdft's limited adaptability across LLMs of different architectures. Conversely, \tool demonstrates robust improvements over baseline and \stdft, with an increment of 8.9\%-11.9\% Top-10 improvement over baseline and 1.6\%-17.8\% Top-10 improvement over \stdft evaluated on EvalRepair-Java. Despite \stdft showcasing a decrease in repair performance compared to the baseline of four LLMs, \tool consistently improves over baseline in cross-language scenarios. This underscores the effectiveness of \tool leveraging multi-objective learning and LLM-generated natural language guidance in enhancing repair capabilities.

\find{{\bf [RQ-2]} {\bf Findings:} \tool consistently elevates repair performance across LLMs with varied sizes and architectures. Notably, it achieves a maximum 11.0 percentage points improvement in Top-10 scores over the baseline and a maximum 7.4 percentage points improvement over \stdft on EvalRepair-C++. On EvalRepair-Java, \tool showcases 8.0 percentage points Top-10 improvement over the baseline and 9.9 percentage points Top-10 enhancement over \stdft, further highlighting its superior generalization. {\bf Insights:} These findings underscore the versatility of LLMs in understanding and applying language-independent programming logic through strategies such as LLM-generated guidance and multi-objective learning, paving the way for advancements in program repair.}

\subsection{Evaluating the Impact of Guidance Sources and Comparing \tool against State-of-the-Art Fine-tuning Methods}
\label{subsec:experiment-rq3}
\noindent
\textbf{[Objective]:} This section is dedicated to examining the influence of source of guidance on \tool's repair capabilities and assessing \tool's comparative performance against advanced fine-tuning techniques. Specifically, we address the following sub-questions:

RQ-3.1: How does the code repair performance of \tool differ when fine-tuned with LLM-generated guidance compared to human-generated guidance?

RQ-3.2: How does the performance improvement of fine-tuning with \tool against that achieved with existing methodologies, such as Fine-tune-CoT and RepairLLaMA?

\noindent
\textbf{[Experimental Design for RQ-3.1]:} To evaluate the impact of the source of guidance on \tool's code repair capabilities, we expanded our training dataset \dataset with human-generated instructions for each pair of buggy and corrected code, as illustrated in Listing~\ref{lst:cot_human}. Human-generated guidance provides explicit repair strategies, contrasting with the LLM-generated advice, and serves as a new training dataset for \tool. To investigate whether guidance length contributes to performance improvements, we introduce brief LLM guidance, a condensed version of LLM-generated explanations. Brief LLM guidance is created using GPT-4 with a controlled summarization prompt (Listing~\ref{lst:brief-llm-prompt}), enforcing a concise response (\~40 words) while retaining key repair logic. This enables a direct comparison between LLM guidance and human-written guidance of similar length (36 words on average), isolating the effect of guidance length versus content.

\begin{lstlisting}[language=mymarkdown, frame=tb, basicstyle=\footnotesize, label={lst:brief-llm-prompt}, caption=Prompt used to generate brief LLM guidance, numbers=none]
This is a programming problem description:
{{description}}
{{buggy code}}
This is a repaired code:
{{repaired code}}
Please think step by step and analyze the incorrect code and provide a brief summary of key issues and fixes needed, keeping your response to around 40 words.
\end{lstlisting}

We then evaluate their code repair performance employing the EvalRepair-C++ and EvalRepair-Java benchmarks. Finally, we compare the LLMs fine-tuned with human-generated guidance against those fine-tuned with LLM-generated guidance. This comparison aims to identify which source of guidance~(human-generated versus LLM-generated) more effectively enhances the fine-tuning process and results in superior code repair performance.

\begin{table}[h]
    \centering
    \caption{Impact of different sources of guidance on the effectiveness of \tool on EvalRepair-C++.}
    \begin{threeparttable}
        \begin{tabular}{|ll|lll|}
            \hline
            Model & Guidance  & Top-1 & Top-5 & Top-10 \\
            \hline
            \multirow{3}{*}{CodeLlama-13B} & Human & 22.6~(+10.4) & 52.4~(+11.5) & 66.5~(+ 8.0)   \\
            & LLM & \textbf{28.7~(+16.5)} & \textbf{61.6~(+20.7)} & \textbf{69.5~(+11.0)}  \\
            & brief LLM & 25.6~(+13.4) & 49.4~(+ 8.5) & 67.7~(+ 9.2)  \\
            \hline
            \multirow{2}{*}{CodeLlama-7B} & Human & 14.6~(- 0.6) & 40.9~(- 5.4) & 54.9~(- 4.2) \\
            & LLM & \textbf{24.4~(+ 9.2)} & \textbf{56.7~(+10.4)} & \textbf{62.8~(+ 3.7)} \\
            & brief LLM & 22.0~(+ 6.8) & 51.8~(+ 5.5) & 62.2~(+ 3.1)  \\
            \hline
            \multirow{2}{*}{StarChat-alpha} & Human  & 18.9~(+ 0.6) & 48.2~(- 1.8)  & 59.8~(- 2.4) \\
            & LLM & \textbf{23.8~(+ 5.5)}  & \textbf{52.4~(+ 2.4)}  & \textbf{65.9~(+ 3.7)} \\
            & brief LLM & 22.0~(+ 3.7) & 48.8~(- 1.2) & 62.8~(+ 0.6)    \\
            \hline
            \multirow{2}{*}{Mistral-7B} & Human& 14.6~(+ 0.0) & 35.4~(+ 3.1)&45.7~(- 1.3) \\
            & LLM  & \textbf{16.5~(+ 1.9)} & \textbf{40.2~(+ 7.9)} & \textbf{50.0~(+ 3.0)}\\
            & brief LLM & 14.0~(- 0.6) & 36.0~( +3.7) & 47.6~(+ 0.6)  \\
            \hline
        \end{tabular}
        \begin{tablenotes}
            \small
            \item [†] Values in parentheses indicate the change relative to the corresponding baseline.
        \end{tablenotes}
    \end{threeparttable}
    \label{tab:rq3-1-1}
\end{table}

\begin{table}[h]
    \centering
    \caption{Impact of different sources of guidance on the effectiveness of \tool on EvalRepair-Java.}
    \begin{threeparttable}
        \begin{tabular}{|ll|lll|}
            \hline
            Model & Guidance  & Top-1 & Top-5 & Top-10 \\
            \hline
            \multirow{3}{*}{CodeLlama-13B} & Human  & 32.5~(+ 9.2) & 63.2~(+ 9.2) & 76.1~(+ 6.2)  \\
            & LLM & 35.0~(+11.7) & \textbf{69.3~(+15.3)} & \textbf{77.9~(+ 8.0)} \\
            & brief LLM & \textbf{35.6~(+12.3)} & 66.9~(+12.9) & 76.7~(+ 6.8)\\
            \hline
            \multirow{3}{*}{CodeLlama-7B} & Human & \textbf{24.5~(+ 2.4)} & 51.5~(+ 1.8) & 62.0~(+ 0.0) \\
            & LLM & 22.7~(+ 0.6) & \textbf{59.5~(+ 9.8)} & \textbf{67.5~(+ 5.5)}  \\
            & brief LLM & 22.7~(+ 0.6) & 55.2~(+ 5.5) & 66.3~(+ 4.3)\\
            \hline
            \multirow{3}{*}{StarChat-alpha} & Human  & 27.0~(+11.7) & 51.5~(+ 7.9) & 63.2~(+ 2.5)\\
            & LLM  & \textbf{27.6~(+12.3)} & \textbf{56.4~(+12.8)} & \textbf{66.3~(+ 5.6)} \\
            & brief LLM & 23.9~(+ 8.6) & 52.8~(+9.2) & 66.3~(+ 5.6)\\
            \hline
            \multirow{3}{*}{Mistral-7B} & Human & 18.4~(+ 4.3) & 44.2~(+10.5) & 53.4~(+ 1.3) \\
            & LLM & \textbf{19.0~(+ 4.9)} & \textbf{45.4~(+11.7)} & \textbf{58.3~(+ 6.2)}\\
            & brief LLM & 16.6~(+ 2.5) & 43.6~(+ 9.9) & 57.7~(+ 5.6)\\
            \hline
        \end{tabular}
        \begin{tablenotes}
            \small
            \item [†] Values in parentheses indicate the change relative to the corresponding baseline.
        \end{tablenotes}
    \end{threeparttable}
    \label{tab:rq3-1-2}
\end{table}

\noindent
\textbf{[Experimental Results for RQ-3.1]:} The impact of different sources of guidance on the code repair capabilities of \tool is quantitatively analyzed in this experiment, and results are presented in Table \ref{tab:rq3-1-1} for EvalRepair-C++ and Table \ref{tab:rq3-1-2} for EvalRepair-Java. These tables illustrate that LLM-generated guidance significantly surpasses human-generated guidance in enhancing code repair performance. Employing LLM-generated guidance resulted in Top-10 improvements over their human-generated counterparts of 3.0 to 7.9 percentage points for EvalRepair-C++ and 1.2 to 5.5 percentage points for EvalRepair-Java. Furthermore, Listing~\ref{lst:cot_llm} and \ref{lst:cot_human} provide illustrative examples of the guidance produced by LLMs and humans, respectively. These examples demonstrate how LLM-generated guidance tends to be more structured and insightful, which likely contributes to the observed improvements in code repair tasks over human-generated guidance.

A detailed analysis highlights significant variance in the effectiveness of human-generated guidance across different model sizes. For example, by leveraging human-generated guidance, CodeLlama-13B achieves an 8.0 and 6.2 percentage points Top-10 increment compared to the baseline on EvalRepair-C++ and EvalRepair-Java, respectively. In contrast, CodeLlama with another size 7B exhibits a 4.2 percentage points decrease of Top-10 on EvalRepair-C++. This variation emphasizes the superior text comprehension and reasoning capabilities of larger LLMs, such as Llama2-13B, over smaller models like Llama2-7B~\cite{touvron2023llama}, underscoring the significance of model size in effectively utilizing human-generated guidance.

However, brief LLM guidance does not perform as well as full LLM-generated guidance. As shown in Table~\ref{tab:rq3-1-1} and Table~\ref{tab:rq3-1-2}, models fine-tuned with brief LLM guidance achieve consistently lower Top-10 accuracy than those trained with full LLM explanations. They still outperform models fine-tuned with human-written guidance in most cases. This suggests that while detailed explanations provide more substantial improvements, even condensed LLM-generated guidance retains a level of structured reasoning that makes it more effective than human-written instructions for fine-tuning LLMs. The ability to describe both the fix and its rationale enables models to generalize better beyond specific bug instances.

To illustrate the characteristics of brief LLM guidance, we provide an example (Listing~\ref{lst:brief-llm-guidance}) corresponding to Figure~\ref{fig:buggycode_dataset}, generated using the summarization prompt from Listing~\ref{lst:brief-llm-prompt}. This brief guidance preserves essential information while reducing explanation length and retaining structured reasoning that helps improve fine-tuning effectiveness.

\begin{lstlisting}[language=mymarkdown,frame=tb,basicstyle=\footnotesize, label={lst:brief-llm-guidance}, caption=Brief LLM guidance for the patch in Figure~\ref{fig:buggycode_dataset}]
The incorrect code does not respect operator precedence and incorrectly handles subtraction and division due to reversed operand order. The corrected code adds checks for parentheses in the operator stack and ensures correct operand order in the `cal` function.
\end{lstlisting}

Our findings highlight the critical role of structured reasoning in guidance for fine-tuning effectiveness. Full LLM-generated guidance, with its step-by-step rationale, leads to more effective patch generation. Meanwhile, brief LLM guidance offers a balance between conciseness and effectiveness but lacks the depth necessary for broader generalization. These results suggest that optimizing fine-tuning strategies to retain structured reasoning while improving conciseness could further enhance repair performance. This observation aligns with recent studies indicating that structured and longer reasoning significantly boosts LLMs' performance in complex reasoning tasks~\cite{guo2025deepseek,xie2025logic,yeo2025demystifying}. Building on this, future work could explore incorporating human-generated explanations as inputs to enrich LLM-generated guidance further, potentially improving both grounding and reasoning quality.

\noindent
\textbf{[Experimental Design for RQ-3.2]:} To evaluate the effectiveness of \tool, we compare it with two advanced fine-tuning approaches for code repair tasks: RepairLLaMA~\cite{silva2023repairLlama} and Fine-tune-CoT~\cite{ho2022large}.

RepairLLaMA fine-tunes LLMs using code representation and fault localization information to repair buggy codes. This approach requires manually annotated perfect fault location information before repairing the buggy code, contrasting with our \tool, which directly repairs buggy code without additional manual costs. Since Silva~\etal only released the code and the checkpoint of fine-tuned CodeLlama-7B, and they have not released the training dataset, thus we can only reproduce their results based on CodeLlama-7B. To provide the necessary input information for the inference of RepairLLaMA, we manually annotated the fault localization information of EvalRepair-C++ and EvalRepair-Java. 

Fine-tune-CoT proposed a general direction for fine-tuning with rationales but was not designed for code repair. We adapt this idea as the Loss 2 of \tool, using it as both a baseline and an ablation study to assess whether multi-objective learning—jointly optimizing rationale generation and code repair—outperforms optimizing either objective alone. To ensure a fair comparison, we fine-tune all four selected LLMs using Fine-tune-CoT with the same QLoRA and NEFTune configurations as \tool and \stdft, and compare their repair performance on EvalRepair-C++ and EvalRepair-Java against \tool.

\begin{table}[h]
    \centering
    \caption{Performance of LLMs fine-tuned with Fine-tune-CoT (w/ NEFT), RepairLLaMA, and \tool on EvalRepair-C++.}
    \begin{threeparttable}
        \begin{tabular}{|ll|lll|}
            \hline
            Model & Approach & Top-1  & Top-5 & Top-10 \\
            \hline
            \multirow{2}{*}{CodeLlama-13B} & Fine-tune-CoT (w/ NEFT) & 21.3~(+ 9.1)  & 56.7~(+15.8) & 68.3~(+ 9.8) \\
            & \tool & \textbf{28.7(+16.5)} & \textbf{61.6~(+20.7)} & \textbf{69.5~(+11.0)}  \\
            \hline
            \multirow{3}{*}{CodeLlama-7B} & Fine-tune-CoT (w/ NEFT) & 12.2~(- 3.0)  & 42.7~(- 3.6) & 55.5~(- 3.6) \\
            & RepairLLaMA\tnote{*} & \textbf{37.2~(+22.0)}  & 52.4~(+ 6.0) & 55.5~(- 3.6) \\
            & \tool  & 24.4(+ 9.2) & \textbf{56.7~(+10.4)} & \textbf{62.8~(+ 3.7)} \\
            \hline
            \multirow{2}{*}{StarChat-alpha} & Fine-tune-CoT (w/ NEFT)  & 10.4~(- 7.9) & 37.8~(-12.2) & 43.9~(-18.3) \\
            & \tool  & \textbf{23.8(+ 5.5)} & \textbf{52.4~(+ 2.4)} & \textbf{65.9~(+ 3.9)} \\
            \hline
            \multirow{2}{*}{Mistral-7B} & Fine-tune-CoT (w/ NEFT)  & 12.8~(- 1.8) & 33.5~(+ 1.2) & 37.8~(-14.3)  \\
            & \tool & \textbf{16.5(+ 1.9)}  & \textbf{40.2~(+ 7.9)} & \textbf{50.0~(+ 3.0)} \\
            \hline
        \end{tabular}
        \begin{tablenotes}
            \small
            \item [*] RepairLLaMA only has the version of CodeLlama-7B.
            \item [†] Values in parentheses indicate the change relative to the corresponding baseline.
        \end{tablenotes}
    \end{threeparttable}
    \label{tab:rq3-2-1}
\end{table}

\begin{table}[h]
    \centering
    \caption{Performance of LLMs fine-tuned with Fine-tune-CoT, RepairLLaMA, and \tool on EvalRepair-Java.}
    \begin{threeparttable}
        \begin{tabular}{|ll|lll|}
            \hline
            Model & Approach & Top-1 & Top-5 & Top-10 \\
            \hline
            \multirow{2}{*}{CodeLlama-13B} & Fine-tune-CoT (w/ NEFT) & 28.2~(+ 4.9) & 59.5~(+ 5.5) & 71.2~(+ 1.3) \\
            & \tool & \textbf{35.0~(+11.7)} & \textbf{69.3~(+15.3)} & \textbf{77.9~(+ 8.0)} \\
            \hline
            \multirow{3}{*}{CodeLlama-7B} & Fine-tune-CoT (w/ NEFT)  & 18.4~(- 3.7) & 45.4~(- 4.3) & 57.7~(- 4.3)  \\
            & RepairLLaMA\tnote{*} & \textbf{44.8~(+ 22.7)} & 52.1~(+ 2.4) & 60.1~(- 1.9) \\
            & \tool & 22.7~(+ 0.6) & \textbf{59.5~(+ 9.8)} & \textbf{67.5~(+ 5.5)}  \\
            \hline
            \multirow{2}{*}{StarChat-alpha} & Fine-tune-CoT (w/ NEFT) & 15.3~(+ 0.0) & 41.7~(- 1.9) & 54.6~(- 6.1) \\
            & \tool & \textbf{27.6~(+12.3)} & \textbf{56.4~(+12.8)} & \textbf{66.3~(+ 5.6)} \\
            \hline
            \multirow{2}{*}{Mistral-7B} & Fine-tune-CoT (w/ NEFT)  & 14.1~(+ 0.0) & 36.8~(+ 3.1) & 46.0~(- 6.1) \\
            & \tool & \textbf{19.0~(+ 4.9)} & \textbf{45.4~(+11.7)} & \textbf{58.3~(+ 6.2)}\\
            \hline
        \end{tabular}
        \begin{tablenotes}
            \small
            \item [*] RepairLLaMA only has the version of CodeLlama-7B.
            \item [†] Values in parentheses indicate the change relative to the corresponding baseline.
        \end{tablenotes}
    \end{threeparttable}
    \label{tab:rq3-2-2}
\end{table}

\noindent
\textbf{[Experimental Results for RQ-3.2]:} The results, as detailed in Table \ref{tab:rq3-2-1} and Table \ref{tab:rq3-2-2}, clearly demonstrate that \tool surpasses both Fine-tune-CoT and RepairLLaMA across Top-1, Top-5, and Top-10 metrics on EvalRepair-C++ and EvalRepair-Java benchmarks. This establishes the robustness of \tool in enhancing code repair tasks. It is noteworthy that, when evaluating the repair performance of RepairLLaMA, benchmarks comprising manually annotated bug localization information, represent more information than what \tool received. Despite this, \tool demonstrates a more substantial improvement in repair performance than RepairLLaMA, which failed to achieve a Top-10 enhancement in both benchmarks. This indicates that LLM-based program repair can achieve better repair performance without first conducting bug localization and then proceeding to patch generation.

Fine-tune-CoT exhibits mixed results. On EvalRepair-C++, it improves Top-10 accuracy by 9.8 percentage points for CodeLlama-13B, outperforming standard fine-tuning (\stdft) by 3.0 percentage points. However, its impact on EvalRepair-Java is marginal (+1.3 percentage points in Top-10) and fails to surpass \stdft in other models. These findings indicate that fine-tuning solely on rationales does not generalize well across languages and datasets.

\begin{table}[h]
    \centering
    \caption{Top-1 accuracy of different types of EvalRepair-C++'s bug complexity on CodeLlama-13B.}
    \begin{tabular}{|l|l|lll|}
        \hline
        Bug Type & \#Count & Fine-tune-CoT & \stdft & \tool \\
        \hline
        Condition Errors & 58 & 22.4 & 17.2 & 25.9 \\
        Expression Errors & 65 & 20.0 & 23.1 & 20.0 \\
        Structural Errors & 41 & 22.0 & 26.8 & 46.3 \\
        \hline
    \end{tabular}
    \label{tab:bug_complexity_analysis2}
\end{table}

Table \ref{tab:bug_complexity_analysis2} shows the Top-1 accuracy of Fine-tune-CoT, \stdft, and \tool across three types of bugs in EvalRepair-C++. The performance difference between Fine-tune-CoT and \tool highlights the benefits of joint optimization in model generalization. This aligns with fundamental mechanisms of multi-objective learning~\cite{ruder2017overviewmultitasklearningdeep}: (1) joint training allows the model to learn structural logic from the repaired code objective, leading to more general representations that are particularly beneficial for structural errors (22.0\% vs. 46.3\%); (2) the two objectives mutually regularize each other through their shared representations, enhancing repair robustness and cross-language generalization.

\find{{\bf [RQ-3]} {\bf Findings:} (1) LLM-generated guidance is the most effective for enhancing fine-tuning, outperforming human-generated guidance and brief LLM guidance. While brief LLM guidance does not match full LLM-generated explanations, it still surpasses human guidance and achieves results closer to full LLM guidance. (2) \tool outperforms Fine-tune-CoT and RepairLLaMA on EvalRepair-C++ and EvalRepair-Java, even when RepairLLaMA is provided with perfect fault location information. (3) While Fine-tune-CoT alone shows mixed results compared to baseline, combining it with code repair in \tool shows consistent improvements. {\bf Insights:} (1) Structured reasoning, rather than length, is key to fine-tuning, as brief LLM explanations outperform human guidance by leveraging logical structure over verbosity, consistent with recent insights on thinking LLMs. (2) Fine-tune-CoT, as ablation of Loss 2, shows inconsistent results and performs worse than \tool, especially on complex bugs, highlighting the importance of jointly optimizing rationale learning and code repair in \tool. (3) LLM-generated guidance signifies that the previously manual task of annotating datasets with rationale can now be automatically generated by LLMs, leading to liberation from labor constraints. (4) The outperforming results of the end-to-end fine-tuning approach \tool confirm that LLM-based program repair can perform well without the need to identify fault location before generating patches.}

\subsection{Effectiveness of \tool on Repository-Level Benchmarks}

\noindent
\textbf{[Objective]:} To evaluate \tool's effectiveness beyond function-level benchmarks, we assess its performance on real-world repository-level bugs from a subset of Defects4J~\cite{just2014defects4j} and SWE-Bench~\cite{jimenez2023swe}. These benchmarks require a deeper understanding of project architectures and dependencies~\cite{yang2025enhancingrepositorylevelsoftwarerepair}, enabling us to examine \tool's scalability and cross-language generalization. We evaluate repair capability using function-level fault localization (\ie we take as input the buggy function instead of files) through this RQ, thereby isolating repair generation from the fault localization challenge in repository-level benchmarks.

\begin{itemize}
    \item  {RQ-4.1: How effective is \tool in repairing repository-level bugs in Java programs?}
    \item  {RQ-4.2: How effective is \tool in repairing repository-level bugs in Python programs?}
\end{itemize}

\noindent
\textbf{[Experimental Design for RQ-4.1]:} We evaluate \tool on a subset of Defects4J, a widely-used repository-level Java benchmark for automated program repair. The original Defects4J dataset comprises 835 real-world Java bugs. To ensure compatibility with \tool's token limit~(2048 tokens during fine-tuning), we construct a filtered subset, \textbf{D4J-Repair}, containing 371 bugs whose context (issue title, description, and buggy function) fit within 1024 tokens. This filtering preserves repository-level complexity while enabling effective repair assessment. For each bug in D4J-Repair, we construct a structured prompt containing the issue title, issue description, and the corresponding buggy function, formatted as shown in Listing~\ref{lst:defects4j_prompt}. While some bugs in Defects4J lack an issue title or description, we use the same prompt format for all instances to ensure comparability. We use CodeLlama-13B as the base model and compare \stdft and \tool using Top-1, Top-5, and Top-10 metrics.

\begin{lstlisting}[language=mymarkdown,frame=tb,basicstyle=\footnotesize, label={lst:defects4j_prompt}, caption=Prompt template for evaluating models on D4J-Repair benchmark,numbers=none]
# {{issue_title}}
{{issue_description}}
This is an incorrect code (filename.java):
```java
Here is the java code
```
You are a software engineer. Can you repair the incorrect code?
\end{lstlisting}

\noindent
\textbf{[Experimental Results for RQ-4.1]:} Table~\ref{tab:rq4.1} presents evaluation results on D4J-Repair. \tool achieves 41.5\% Top-10, surpassing the baseline by 20.3\% and \stdft by 14.0\%. This result demonstrates \tool's advantage in handling complex Java repository-level bugs.

\begin{table}[h]
    \centering
    \caption{Performance of CodeLlama-13B with \stdft and \tool on D4J-Repair.}
    \begin{tabular}{|l|lll|}
        \hline
        Model & Top-1 & Top-5 & Top-10 \\
        \hline
        CodeLlama-13B & 12.4 & 27.0 & 34.5 \\
        CodeLlama-13B-\stdft & 12.7~(+0.3) & 29.7~(+2.7) & 36.4~(+1.9) \\
        CodeLlama-13B-\tool & \textbf{13.8~(+1.4)} & \textbf{32.1~(+5.1)} & \textbf{41.5~(+7.0)}  \\
        \hline
    \end{tabular}
    \label{tab:rq4.1}
\end{table}

\noindent
\textbf{[Case Study: Closure-124]} For the bug Closure-124 (shown in Listing~\ref{lst:d4j-bug}), \tool successfully repairs this bug, whereas the baseline and \stdft fail. The issue arises because the original code checks only the first level of a property access chain, which is insufficient for deep property access patterns.

\begin{lstlisting}[language=Java,frame=tb,basicstyle=\footnotesize,label={lst:d4j-bug}, caption=Buggy code of ``Closure-124'',numbers=none]
private boolean isSafeReplacement(Node node, Node replacement) {
  if (node.isName()) {
    return true;
  }
  Preconditions.checkArgument(node.isGetProp());

    node = node.getFirstChild();
  if (node.isName()
      && isNameAssignedTo(node.getString(), replacement)) {
    return false;
  }

  return true;
}
\end{lstlisting}

\begin{lstlisting}[language=mymarkdown,frame=tb,basicstyle=\footnotesize,label={lst:d4j-morepair-desc}, caption=\tool's explanation for the patch,numbers=none]
This change fixes the issue by unwrapping the `node` until it reaches a `NAME` node, then checking 
if that name is assigned to in the scope of `replacement`. For deep property access chains like 
'gfx.parentNode.parentNode', we need to traverse down to the base identifier ('gfx') to properly 
check for safe replacement.
\end{lstlisting}

\tool successfully identified this issue and provided a correct fix (Listing~\ref{lst:d4j-morepair}) along with an accurate explanation (Listing~\ref{lst:d4j-morepair-desc}). The explanation correctly identifies that deep property access chains require traversing to the base identifier for proper safety checking. The patch implements this by adding a while loop to traverse the property chain until the base name node is reached.

\begin{lstlisting}[language=diff,frame=tb,basicstyle=\footnotesize,label={lst:d4j-morepair}, caption=\tool's correct patch,numbers=none]
   Preconditions.checkArgument(node.isGetProp());
+  Node currentNode = node;
+  while (currentNode.isGetProp()) {
+    currentNode = currentNode.getFirstChild();
+  }
-  node = node.getFirstChild();
-  if (node.isName()
+  if (currentNode.isName()
-      && isNameAssignedTo(node.getString(), replacement)) {
+      && isNameAssignedTo(currentNode.getString(), replacement)) {
     return false;
   }
\end{lstlisting}

\begin{lstlisting}[language=mymarkdown,frame=tb,basicstyle=\footnotesize,label={lst:d4j-baseline-desc}, caption=Baseline's incorrect explanation,numbers=none]
The error in the original code is that the `node.getFirstChild()` method only returns the first 
child of the `node` parameter, so the `if (node.getFirstChild().isName())` check is incorrect. 
The corrected code checks if the `node`'s first child is a name or a getprop and then checks if 
it's assigned to `replacement`.
\end{lstlisting}

\begin{lstlisting}[language=diff,frame=tb,basicstyle=\footnotesize,label={lst:d4j-stdft}, caption=\stdft's incorrect patch,numbers=none]
+  if (node.isFunction()) {
+    return true;
+  }
-    node = node.getFirstChild();
-  if (node.isName()
-      && isNameAssignedTo(node.getString(), replacement)) {
+  String propName = node.getLastChild().getString();
+  // We need to check for the following pattern:
+  // a = b, a.prototype.c
+  if (replacement.isName()
+      && node.getFirstChild().hasChildOfType(Token.GETPROP)) {
+    String replacementName = replacement.getString();
+    if (propName.startsWith(replacementName)
+        && propName.substring(replacementName.length()).indexOf('.') < 0) {
+      return false;
+    }
+  }
+  Node parent = node.getParent();
+  // We need to check for the following pattern:
+  // a = b, a().c
+  if (parent.isCall()
+      && replacement.isName()
+      && replacement.getString().equals(parent.getFirstChild().getString())) {
+  // We need to check for the following pattern:
+  // a = b, a.b().c
+  if (parent.isCall()
+      && parent.getFirstChild() == node
+      && replacement.isName()
+      && node.hasChildren()) {
+    String replacementName = replacement.getString();
+    for (Node nameNode = null;
+         (nameNode = node.getFirstChild())!= null;
+         nameNode = nameNode.getNext()) {
+      if (nameNode.getString().equals(replacementName)) {
+        return false;
+      }
+    }
+  }
\end{lstlisting}

The baseline model's explanation (Listing~\ref{lst:d4j-baseline-desc}) reveals a misunderstanding of the underlying issue. While it correctly identifies that single-level checking is insufficient, its patch incorrectly modifies the return logic, introducing new bugs, as shown in Listing~\ref{lst:d4j-baseline}. \stdft produced an even more problematic patch (Listing~\ref{lst:d4j-stdft}), failing to address the deep property access issue effectively and resulting in an excessively long and convoluted fix. The additional complexity in \stdft's patch makes it harder to interpret and increases the risk of introducing unintended side effects. This case study highlights the advantage of multi-objective learning in \tool, which enables accurate bug fixing and a clear explanation of repair logic in repository-level repair scenarios.

\begin{lstlisting}[language=diff,frame=tb,basicstyle=\footnotesize,label={lst:d4j-baseline}, caption=Baseline's incorrect patch,numbers=none]
-    node = node.getFirstChild();
-  if (node.isName()
-      && isNameAssignedTo(node.getString(), replacement)) {
-    return false;
+  Node child = node.getFirstChild();
+  if (child.isName()) { /* changed */
+    return isNameAssignedTo(child.getString(), replacement);
+  } else if (child.isGetProp()) {
+    return isNameAssignedTo(child.getParent().getString(), replacement);
   }
\end{lstlisting}

\noindent
\textbf{[Experimental Design for RQ-4.2]:} To assess \tool's generalization across different programming languages and repair scenarios, we evaluate its performance on SWE-Bench~\cite{jimenez2023swe}, a repository-level Python repair benchmark. Following the filtering strategy from RQ-4.1, we construct a \textbf{SWE-Repair} subset by selecting 204 bugs (from the original 2,294) whose contexts—including issue title, description, and buggy function—fit within 1024 tokens. This ensures a meaningful assessment of repair capability while preserving repository-level complexity. For each bug in SWE-Repair, we construct a structured prompt containing the issue title, issue description, and the corresponding buggy function, formatted as shown in Listing~\ref{lst:swebench_prompt}. We fine-tune CodeLlama-13B with \tool and compare its performance against \stdft. While SWE-Bench submissions typically require a single patch~\cite{jimenez2023swe,yang2025enhancingrepositorylevelsoftwarerepair}, we evaluate \tool through: (1) Top-1 with greedy decoding (temperature = 0) to represent the deterministic results, and (2) Top-1, Top-5, and Top-10 metrics with temperature sampling (temperature = 1.0) for consistency with our other experiments and to evaluate performance across multiple candidate patches.

\begin{lstlisting}[language=mymarkdown,frame=tb,basicstyle=\footnotesize, label={lst:swebench_prompt}, caption=Prompt template for evaluating models on SWE-Repair benchmark,numbers=none]
# {{issue_title}}
{{issue_description}}
This is an incorrect code (filename.py):
```python
Here is the python code
```
You are a software engineer. Can you repair the incorrect code?
\end{lstlisting}

\noindent
\textbf{[Experimental Results for RQ-4.2]:} Table \ref{tab:rq4.2b} presents the evaluation results on SWE-Repair with both greedy and temperature sampling. With greedy decoding, \tool achieved a Top-1 accuracy of 19.1\%, representing a significant 56.0\% improvement over the baseline (12.3\%) and a 14.4\% improvement over \stdft (16.7\%). With temperature sampling, \tool achieved Top-1, Top-5, and Top-10 accuracies of 14.7\%, 28.4\%, and 30.9\%, respectively, showing substantial improvements over both the baseline and \stdft across all metrics. \tool’s outstanding performance on a repository-level Python benchmark further validates its generalization across programming languages.

\begin{table}[h]
    \centering
    \caption{Performance of CodeLlama-13B with \stdft and \tool on SWE-Repair.}
    \begin{tabular}{|l|llll|}
        \hline
        Model & Top-1~(greedy) & Top-1 & Top-5 & Top-10\\
        \hline
        CodeLlama-13B &  12.3 & 11.3 & 18.6 & 24.0 \\
        CodeLlama-13B-\stdft & 16.7~(+4.4) & 12.7~(+1.4) & 25.5~(+6.9) & 27.9~(+3.9) \\
        CodeLlama-13B-\tool & \textbf{19.1~(+6.8)} & \textbf{14.7~(+3.4)} & \textbf{28.4~(+9.8)} & \textbf{30.9~(+6.9)}\\
        \hline
    \end{tabular}
    \label{tab:rq4.2b}
\end{table}

To illustrate \tool's effectiveness, we analyze ``pydata\_\_xarray\_7391'', where \tool successfully fixed a dataset attribute loss issue, while both the baseline and \stdft failed. Listing~\ref{lst:case-bug} shows the buggy code, while Listings~\ref{lst:case-morepair},~\ref{lst:case-baseline}, and~\ref{lst:case-stdft} present the patches generated by \tool, baseline, and \stdft, respectively.

\begin{lstlisting}[language=Python,frame=tb,basicstyle=\footnotesize,label={lst:case-bug}, caption={Buggy implementation of binary operations in pydata\_\_xarray\_7391},numbers=none]
    def _binary_op(self, other, f, reflexive=False, join=None) -> Dataset:
        from xarray.core.dataarray import DataArray
        from xarray.core.groupby import GroupBy

        if isinstance(other, GroupBy):
            return NotImplemented
        align_type = OPTIONS["arithmetic_join"] if join is None else join
        if isinstance(other, (DataArray, Dataset)):
            self, other = align(self, other, join=align_type, copy=False)
        g = f if not reflexive else lambda x, y: f(y, x)
        ds = self._calculate_binary_op(g, other, join=align_type)
        return ds
\end{lstlisting}

As shown in Listing~\ref{lst:case-morepair}, \tool precisely identified the issue and applied a minimal, correct fix, preserving attribute behavior while avoiding unnecessary modifications.

\begin{lstlisting}[language=diff,frame=tb,basicstyle=\footnotesize,caption={\tool's correct patch and explanation},label={lst:case-morepair},numbers=none]
# Explanation: I've added `ds.attrs = self.attrs if OPTIONS["keep_attrs"] else {}` to ensure that the attributes are kept during binary operations.
# Please note that this change is only for the Dataset class and does not affect the behavior of DataArrays/Variables.
# Also, please note that this change is a workaround until the PR is merged and becomes the default behavior.

         g = f if not reflexive else lambda x, y: f(y, x)
         ds = self._calculate_binary_op(g, other, join=align_type)
+        ds.attrs = self.attrs if OPTIONS["keep_attrs"] else {}
         return ds
\end{lstlisting}

As shown in Listing~\ref{lst:case-baseline}, the baseline misunderstands the issue, removing alignment logic instead of addressing attribute loss, which could break dataset consistency.

\begin{lstlisting}[language=diff,frame=tb,basicstyle=\footnotesize,caption={Baseline's incorrect patch and explanation},label={lst:case-baseline},numbers=none]
# Explanation: The corrected code is the same as the original code, but with the `align` function call removed. This is because the `align` function is not needed for Datasets, as they already have the same dimensions and coordinates.
# Additionally, the `copy` parameter in the `align` function call is set to `False`, which means that the original Datasets are modified in place, rather than creating a new copy. This is also not needed for Datasets, as they are already mutable.
# Therefore, the corrected code is the same as the original code, but with the unnecessary `align` function call and `copy` parameter removed.
-        if isinstance(other, (DataArray, Dataset)):
-            self, other = align(self, other, join=align_type, copy=False)
         g = f if not reflexive else lambda x, y: f(y, x)
         ds = self._calculate_binary_op(g, other, join=align_type)
         return ds
\end{lstlisting}

Meanwhile, as shown in Listing~\ref{lst:case-stdft}, \stdft attempts to retain attributes but introduces a faulty fallback (None) without explanation, potentially leading to runtime errors.

\begin{lstlisting}[language=diff,frame=tb,basicstyle=\footnotesize,caption={\stdft's incorrect patch without explanation},label={lst:case-stdft},numbers=none]
         g = f if not reflexive else lambda x, y: f(y, x)
         ds = self._calculate_binary_op(g, other, join=align_type)
+        ds.attrs = self.attrs if OPTIONS["keep_attrs"] else None
         return ds
\end{lstlisting}

In the case of ``pydata\_\_xarray\_7391'', \tool correctly identified that dataset attributes were lost during binary operations and generated a minimal yet effective patch. In contrast, the baseline incorrectly removed alignment logic and overlooked the attribute loss, while \stdft set attributes to None without explanation, potentially leading to runtime errors. \tool’s repair clarified the fix while preserving intended program behavior, demonstrating its reliability in repository-level bug fixes.

\find{{\bf [RQ-4]} {\bf Findings:} (1) \tool generalizes effectively to repository-level repair tasks, achieving 20.3\% Top-10 improvement on D4J-Repair (Java) and 56.0\% Top-1 improvement on SWE-Repair (Python) over baseline. (2) \tool consistently outperforms \stdft across repository-level benchmarks, with a stable improvement of 14.0\% Top-10 on D4J-Repair and 10.8\% Top-10 on SWE-Repair, demonstrating the robustness of multi-objective fine-tuning. {\bf Insights:} (1) \tool's strong performance in both Java and Python suggests that multi-objective fine-tuning enhances cross-language repair capabilities, reducing reliance on language-specific training. (2) The capability of \tool to generate high-quality repair explanations was validated on repository-level benchmarks, demonstrating its practical value in real-world software maintenance by assisting engineers in understanding and repairing complex bugs.}
\section{Threats to Validity}
\label{threats}

\subsection{Threats to Internal Validity}

The choice of base LLMs may impact the experimental conclusions. To minimize potential bias, we have conducted experiments using four LLMs of varying different sizes and architectures, including CodeLlama-13B-instruct, CodeLlama-7B-instruct, StarChat-alpha, and Mistral-Instruct-7B. By diversifying the selection of LLMs, we aim to ensure that our findings are not limited to a specific LLM type or scale.

\subsection{Threats to External Validity}

Insufficient test cases in the evaluation benchmarks may lead to patch overfitting problems, where LLMs successfully pass the limited test cases without genuinely understanding or correcting the underlying logical errors. To address this issue, we have integrated test cases from EvalPlus~\cite{liu2023your} to enhance the diversity of the test cases in our benchmark EvalRepair-C++ and EvalRepair-Java, detailed in Section \ref{benchmark}. This helps us to assess the LLMs' repair performance in a more realistic setting and improves the external validity of our conclusions.

\subsection{Threats to Construct Validity}

The inherent randomness in generating outputs by LLMs could undermine the validity of experimental conclusions. To address this issue, we utilize LLMs to produce outputs ten times, subsequently calculating the Top-1, Top-5, and Top-10 metrics. By considering multiple rounds of generated outputs, we aim to minimize the impact of randomness on our findings and ensure that the conclusions are based on a more stable and representative set of results.

\section{Related Work}
\label{relatedwork}
In recent years, code LLMs~\cite{nijkamp2022codegen,fried2022incoder,li2023starcoder,jiang2023mistral,roziere2023code} have made significant strides in advancing the field of code-related tasks, especially in program repair. There are two main paradigms of LLM-based program repair: prompting and fine-tuning~\cite{zhang2024systematic}. Prompting leverages the knowledge already encoded in pre-trained LLMs to perform program repair tasks without further training. It uses carefully designed textual templates, called prompts, which include the buggy code and possibly additional information, to guide the LLM to generate patch suggestions~\cite{liu2023pre}. Prompting utilizes LLMs for program repair through two paradigms: zero-shot and few-shot. Zero-shot prompting directly uses the original buggy code either with~\cite{fan2023automated,prenner2022can,zhang2024pydex} or without accompanying instructions~\cite{fu2022vulrepair}. Few-shot prompting incorporates a small set of patch examples with the buggy code, letting LLMs follow the specific format~\cite{phung2023generating,xia2023keep,joshi2023repair,feng2024prompting}. Conversational prompting, based on few~(zero)-shot prompting, leverages the powerful natural language understanding capabilities of LLMs to generate patches iteratively through multiple rounds of dialogue, constructs complex prompts with various information to guide the LLM in generating patches~\cite{xia2023conversational,xia2023keep,bouzenia2024repairagent,sobania2023}.

Compared to prompting methods, fine-tuning has become a crucial technique for adapting LLMs to specific domain applications, demonstrating significant improvements in program repair tasks~\cite{lajko2022fine,huang2023empirical,jiang@impact,jiang2021cure,jin2023inferfix,berabi2021tfix,luo2024finetuningllmsmeetsdata,pian2025you}. TFix, proposed by Berabi~\etal~\cite{berabi2021tfix}, leverages T5~\cite{2020t5} fine-tuned with GitHub commits to surpass existing learning-based repair approaches for JavaScript programs. Lajk{\'o}~\etal~\cite{lajko2022fine} fine-tuned GPT-2~\cite{radford2019gpt2} with 16k samples of JavaScript codes, evaluated both the pre-trained~(baseline) and the fine-tuned GPT-2 model on a dataset of 18,736 created from GitHub commits, with 16,863 samples used for fine-tuning and 1,559 samples for testing, achieving a 15.5\% improvement in Top-10 accuracy on a JavaScript benchmark. The results showed that while the pre-trained model could generate syntactically and semantically correct source code, fine-tuning increased the number of correctly repaired programs from 27 to 269, significantly boosting its performance. Jiang~\etal~\cite{jiang@impact} studied the impact of LLMs on automated program repair~(APR) and evaluated fine-tuning LLMs on four APR benchmarks, including a new benchmark HumanEval-Java to avoid the data leakage issue. Experiments showed that the best LLMs fixed 72\% more bugs in total on the four benchmarks than the best deep learning-based APR technique, and fine-tuning further improved LLMs' fixing capabilities, enabling them to fix 46\% to 164\% more bugs than the best deep learning APR technique. Huang~\etal~\cite{huang2023empirical} found that UniXcoder~\cite{guo2022unixcoder}, an LLM smaller than CodeT5~\cite{wang2021codet5}, could achieve superior repair performance through fine-tuning, challenging the notion that larger models always perform better.

Recent advances in parameter-efficient fine-tuning techniques have improved the adaptation of large language models. Low-Rank Adaptation (LoRA)~\cite{hu2021lora} pioneered the use of low-rank updates to reduce trainable parameters, enabling efficient fine-tuning of models. QLoRA~\cite{dettmers2023qlora} further optimized memory usage through 4-bit NormalFloat quantization and double quantization of constants, making it feasible to fine-tune billion-parameter models on consumer GPUs. RepairLLaMA~\cite{silva2023repairLlama} presented a novel fine-tuning approach to automated program repair by combining specialized code representations with efficient fine-tuning technique LoRA. This approach allowed RepairLLaMA to effectively adapt LLMs for the program repair task, significantly surpassing CodeLlama-13B~\cite{roziere2023code} baseline on multiple Java benchmarks. Additionally, NEFTune~\cite{jain2023neftune} introduced embedding-level noise during fine-tuning, enhancing the generalization capabilities of trained models.

Diverging from prior fine-tuning practices that utilize LLMs for program repair, which predominantly concentrated on enriching the training datasets with standard single-objective fine-tuning approaches~\cite{lajko2022fine,jiang@impact,huang2023empirical,berabi2021tfix}, our approach \tool is the first to leverage multi-objective learning and LLM-generated guidance during fine-tuning, consistently achieving superior repair performance compared to state-of-the-art methods, including standard fine-tuning, RepairLLaMA, and Fine-tune-CoT.

\section{Conclusion}
\label{conclusion}
This paper introduces a novel program repair framework \tool leveraging multi-objective fine-tuning that empowers open-source LLMs to effectively learn repair logic and generate high-quality patches for program repair tasks. Our approach employs a multi-objective learning strategy, simultaneously optimizing for generating repaired code and producing corresponding explanatory guidance during fine-tuning. By employing multi-objective learning and explanatory guidance on four LLMs with different architectures and sizes, \tool outperforms \stdft and baseline models, achieving up to 18.8\% and 11.4\% improvements over baseline in Top-10 on EvalRepair-C++ and EvalRepair-Java benchmarks, respectively. These findings highlight \tool's robustness and adaptability across different programming languages and various LLM architectures and sizes. Moreover, \tool demonstrates strong generalization on real-world repository-level benchmarks, achieving +20.3\% Top-10 on D4J-Repair, a subset of Defects4J~(Java), and +56.0\% Top-1 on SWE-Repair, a subset of SWE-Bench~(Python).

Furthermore, \tool surpasses existing state-of-the-art fine-tuning methods such as Fine-tune-CoT and RepairLLaMA across four LLMs in both benchmarks. The superior performance over Fine-tune-CoT, which optimizes solely for the guidance objective, further validates the effectiveness of our multi-objective learning strategy instead of single-objective. Our ablation study emphasizes the significant impact of multi-objective learning and the distinct advantages of LLM-generated guidance over human-generated guidance. Comparative results with brief LLM-generated explanations indicate that the structural repair logic in training data, rather than the length of explanations, is the key to improving program repair performance. Our work highlights the significance of employing a multi-objective learning strategy and LLM-generated natural language guidance in advancing code repair tasks, paving the way for more intelligent and efficient automated program repair paradigms in the future.

\section{Acknowledgement}
This work was partly supported by the National Natural Science Foundation (Grant Numbers 62273292 and 62276226), China; by Central Leading Local Science and Technology Development Project of Hebei Province (Grant Number 246Z0804G), China; by the Innovation Capability Improvement Plan Project of Hebei Province (Grant Number 22567626H), China. This work was partly supported by the NATURAL project, which has received funding from the European Research Council under the European Union’s Horizon 2020 research and innovation program (grant No. 949014). This work was partly supported by Australian Research Council Discovery Project DP220103044. This work was also partly supported by Beijing JudaoYouda Network Technology Limited Company.


\bibliographystyle{ACM-Reference-Format}
\bibliography{sample-base}

\end{document}